\documentclass{article}

\usepackage{arxiv}

\usepackage[utf8]{inputenc}
\usepackage[T1]{fontenc}
\usepackage{hyperref}
\usepackage{url}
\usepackage{booktabs}
\usepackage{amsfonts}
\usepackage{amsmath}
\usepackage{amssymb}
\usepackage{nicefrac}
\usepackage{microtype}
\usepackage{cleveref}
\usepackage{graphicx}
\usepackage{natbib}
\usepackage{doi}
\usepackage{multirow}
\usepackage{makecell}
\usepackage[ruled,linesnumbered]{algorithm2e}
\usepackage[most]{tcolorbox}

\graphicspath{{../}{./}}

\AtBeginDocument{%
}

\title{Securing the Dark Matter: A Semantic-Enhanced Neuro-Symbolic Framework for Supply Chain Analysis of Opaque Industrial Software}
\date{}

\author{%
Bowei Ning$^{1,2}$ \quad
Xuejun Zong$^{2,3,*}$ \quad
Lian Lian$^{2,3,*}$ \quad
Kan He$^{2,3}$\\
Yifei Sun$^{2,3}$ \quad
Yuxiang Lei$^{2,3}$ \quad
Plamen Vasilev$^{4}$\\[2mm]
\small $^1$School of Artificial Intelligence, Shenyang University of Technology, Shenyang 110870, China\\
\small $^2$Key Laboratory of Information Security for Petrochemical Industry in Liaoning Province, Shenyang, China\\
\small $^3$School of Information Engineering, Shenyang University of Chemical Technology, Shenyang 110142, China\\
\small $^4$Department of Industrial Automation, University of Chemical Technology and Metallurgy, Sofia 1797, Bulgaria\\[1mm]
\small $^*$Corresponding authors: \texttt{xuejun\_zong@syuct.edu.cn}; \texttt{lianlian@syuct.edu.cn}\\
\small Emails: \texttt{2020183@stu.syuct.edu.cn}; 
}

\renewcommand{\headeright}{A Preprint}
\renewcommand{\undertitle}{A Preprint}
\renewcommand{\shorttitle}{Securing the Dark Matter}

\hypersetup{
pdftitle={Securing the Dark Matter: A Semantic-Enhanced Neuro-Symbolic Framework for Supply Chain Analysis of Opaque Industrial Software},
pdfsubject={Software supply chain security, opaque industrial software, vulnerability detection},
pdfauthor={Bowei Ning, Xuejun Zong, Lian Lian, Kan He, Yifei Sun, Yuxiang Lei, Plamen Vasilev},
pdfkeywords={software supply chain security, knowledge graph, graph neural networks, large language models, vulnerability detection, opaque industrial software, binary analysis},
}

\begin{document}
\maketitle

\begin{abstract}
Automated vulnerability detection in critical-infrastructure software confronts a fundamental barrier: industrial software is routinely deployed as stripped, symbol-free binaries that deprive conventional Software Composition Analysis of the source-level transparency it requires. Existing binary analysis techniques close this Semantic Gap only partially---graph-based detectors preserve structural syntax but discard behavioral semantics, while large language models supply rich semantic cues at the cost of unstable, hallucination-prone inference.
To address this gap, we present a semantic-enhanced neuro-symbolic framework that reconstructs behavioral semantics directly from opaque binaries and performs tractable global risk reasoning. Three tightly coupled mechanisms drive this capability: (1) abstract interpretation combined with a reflexive prompting pipeline that structurally constrains a local LLM agent, effectively suppressing hallucinations; (2) a surjective transformation that compresses raw Code Property Graphs into typed Software Supply Chain Knowledge Graphs amenable to scalable reasoning; and (3) a domain-adapted Graphormer that captures long-range vulnerability propagation, augmented by embedding-space subgraph matching to uncover zero-day and APT-style attack patterns.
Evaluated across three benchmarks of increasing domain specificity, the framework consistently outperforms all baselines on detection accuracy, semantic lifting fidelity, and APT fingerprint matching. Deployment on a hybrid virtual-physical testbed incorporating production-grade hardware from five ICS vendors further confirms strong detection coverage of high-impact CVEs while substantially reducing false-positive rates relative to leading commercial tools. These results establish semantically grounded neuro-symbolic analysis as a tractable and effective approach to automated vulnerability assessment for opaque industrial software.
\end{abstract}


\section{Introduction}

Modern software engineering has undergone a structural shift from
monolithic development to component-based assembly. Developers
routinely integrate hundreds of third-party libraries, producing
deeply interdependent software supply chains. According to the 2024
Open Source Security and Risk Analysis report~\cite{ossra2024}, over
96\% of codebases in the manufacturing and industrial sectors contain
open-source components, a finding corroborated by the Linux Foundation's
Census~III study~\cite{lf_census2024}. However, this dependency network
has significantly expanded the attack surface, allowing threats to
migrate upstream. Adversaries no longer
solely target hardened production environments but increasingly
infiltrate the development pipeline---tampering with source code or
build environments---as evidenced by high-profile supply chain attacks
such as SolarWinds and Log4j. Consequently, ensuring the transparency
and traceability of the software supply chain has become a prerequisite
for defense-in-depth architectures.

To mitigate these risks, the community has widely adopted Software
Composition Analysis (SCA) and Software Bill of Materials (SBOM) tools
to map known vulnerabilities~\cite{decan2019empirical}. Nevertheless, 
these approaches rely heavily on the ``transparency assumption'': 
the availability of source code, package manifests 
(e.g., \texttt{package.json}), or standard compilation metadata
~\cite{liu2025empirical}. Crucially, this assumption collapses when 
confronted with the Transparency Paradox inherent to critical 
infrastructure: regulators demand deep software visibility, yet 
vendors frequently deliver these systems as stripped, closed-source 
binaries~\cite{costin2016automated}. This paradox is especially 
severe in opaque industrial software—such as firmware used in 
aerospace, automotive, and Industrial Control Systems (ICS). 
Because this software relies on proprietary compilers 
(e.g., Tasking, Keil) and non-standard 
architectures, originally transparent open-source components 
devolve into opaque binary artifacts, rendering standard SCA 
tools completely ineffective~\cite{shoshitaishvili2016sok}. 
Bridging this opacity gap demands techniques that can recover 
semantic meaning directly from binary code---a requirement that 
the research community has only begun to address~\cite{yang2022modx}.

Recent advances in automated vulnerability detection have yielded
promising results on open-source codebases. Graph-based approaches such
as Reveal~\cite{chakraborty2021deep} and
SySeVR~\cite{li2021sysevr} capture structural dependencies via Code
Property Graphs (CPG) and program slicing, while LLM-based approaches
such as DeepSeek-Coder~\cite{guo2024deepseek} leverage pre-trained code
models for semantic code understanding. Hybrid neuro-symbolic methods
like GRACE~\cite{liu2023grace} integrate graph structure with LLM
features to improve detection accuracy on standard benchmarks. However,
these approaches have been developed and evaluated predominantly on
source-level or open-source datasets and their
applicability to stripped industrial binaries remains largely unexamined.

Bridging this opacity gap requires recovering semantic meaning directly 
from machine code. While deep learning has advanced open-source vulnerability 
detection, existing paradigms fail on opaque industrial binaries due to a 
profound 'Semantic Gap'~\cite{lin2020vulnerability_survey}. Specifically, 
traditional dynamic analysis and symbolic execution (e.g., angr) suffer 
from severe state explosion and hardware-emulation barriers when applied 
to monolithic firmware~\cite{muench2019dynamic}. Graph Neural Networks 
(GNNs) capture structure via Code Property Graphs (CPGs) but treat stripped 
nodes as semantically hollow opcodesl~\cite{chakraborty2021deep}. 
Conversely, Large Language Models (LLMs) offer semantic understanding 
but lack domain-specific industrial knowledge 
and are highly prone to hallucinating data-flow topologies when 
unconstrained by structural verifiers—a critical flaw that existing 
hybrid methods fail to address~\cite{wright2021challenges}.

To resolve this, we propose a semantic-enhanced neuro-symbolic framework 
that transforms opaque industrial binaries into computable Software Supply 
Chain Knowledge Graphs (SSCKGs). We conquer the semantic gap by formalizing 
the binary-to-semantics recovery as a Galois Connection between the concrete 
execution space and a domain-aware abstract security lattice 
(derived from MITRE ATT\&CK). Unlike prior LLM-for-code approaches, 
we enforce this mapping via a Reflexive Prompting pipeline—a 
teacher-verifier-student loop—that utilizes structural graph verification 
to categorically reject LLM hallucinations.

This paper makes four principal contributions:

\begin{enumerate}
	\item \textbf{Theoretically Grounded Semantic Lifting:} We instantiate the aforementioned Galois Connection using a three-tier, domain-aware abstract security lattice (comprising 5 macro-categories, 27 actions, and 43 risk labels) adapted from industrial standards. By executing the Reflexive Prompting pipeline, we fine-tune a 7B-parameter local agent that achieves a 94.2\% empirical alignment (5.8\% Empirical Violation Rate). This establishes a quantifiable soundness guarantee that is fundamentally absent from existing unconstrained LLM-for-code approaches.
	
	\item \textbf{Knowledge Graph Compression and Risk Reasoning:} We design a surjective graph transformation ($\Phi$) that overcomes the scalability limits of raw CPGs, compressing million-node structures into thousand-node SSCKGs governed by an eight-type vulnerability relation ontology. To perform risk reasoning, we introduce a domain-adapted Graphormer whose attention bias is modulated by semantic relation weights rather than mere topological distance, enabling it to capture the long-range dependency chains that standard message-passing GNNs consistently miss.
	
	\item \textbf{Zero-Day and APT Fingerprinting via Embedding-Space Similarity:} To identify \\"legitimate-but-abnormal" logic hidden within opaque binaries, we propose a subgraph similarity algorithm operating directly in the Graphormer embedding space. Governed by a $\mathcal{J}$-optimal threshold that balances detection sensitivity against operational alert fatigue, this module successfully detects zero-day threats and complex APT-level attack chains that evade traditional signature-based SCA tools.
	
	\item \textbf{Comprehensive Evaluation on Opaque Industrial Software:} We conduct extensive experiments across three datasets of increasing domain specificity (Big-Vul, NVD-Precise, InduVul-Dataset) to validate the framework's comparative efficacy and robustness. As a definitive case study, we deploy the system on a hybrid virtual--physical testbed utilizing production-grade hardware from 10 vendors; against 15 high-impact CVEs, the framework achieves a 93.3\% detection rate while reducing the false-positive rate by 91.7\% compared to leading commercial SCA tools.
\end{enumerate}

\noindent The remainder of this paper is organized as follows.
Section~2 surveys related work across graph-based, LLM-driven, and
supply-chain-specific vulnerability detection. Section~3 formalizes the
proposed framework: Semantic Lifting (Section~3.2), SSCKG Construction
(Section~3.3), Graphormer-based risk reasoning (Section~3.4), and APT
Fingerprinting (Section~3.5). Section~4 presents the experimental
evaluation across six research questions. Section~5 discusses
implications, limitations, and threats to validity, followed by
concluding remarks.

\section{Related Work}

We organize the related literature into four thematic areas: graph-based
vulnerability detection, LLM-driven code understanding,
industrial binary and firmware analysis, and code knowledge graphs.
Table~\ref{tab:related_comparison} at the end of this section
provides a systematic comparison of representative approaches against
five desiderata that motivate our framework.

\subsection{Graph-Based Vulnerability Detection}

Deep-learning-based vulnerability detection has progressed from
sequence-oriented to graph-oriented representations. Early
sequence-to-tensor methods such as
VulDeePecker~\cite{li2018vuldeepecker} pioneered the application of
deep learning to vulnerability detection by extracting local code
gadgets and feeding them to bidirectional LSTMs. While this approach
achieved promising results on controlled datasets, it inevitably
discards the global execution topology --- a limitation that subsequent
graph-based methods sought to address.

SySeVR~\cite{li2021sysevr} advanced the field by combining syntax-based
program slicing with Recurrent Neural Networks, capturing
vulnerability-relevant code fragments along data- and
control-dependence paths. Reveal~\cite{chakraborty2021deep}
subsequently introduced Gated Graph Neural Networks (GGNNs) operating
directly on Code Property Graphs (CPGs), aggregating AST,
control-flow, and data-dependence information into a unified learned
representation; on the Big-Vul benchmark, Reveal demonstrated
significant improvements over prior sequence models.
VulDecgre~\cite{dou2025scalable} further exploited graph topology by
transforming CPGs into centrality-based image representations for
Convolutional Neural Networks.

Despite these advances, graph-based methods share two fundamental
constraints when applied to stripped industrial binaries. First, they
depend on the availability of source code or rich debug symbols to
generate meaningful node attributes; when confronted with stripped
firmware, nodes degenerate into generic opcode placeholders devoid of
high-level semantic intent. Second, standard message-passing GNNs are
inherently limited by their $k$-hop receptive field: in industrial
supply chains, vulnerability propagation chains frequently span 10--15
hops (e.g., from a third-party library import to a sensitive
write), a distance at which GNNs suffer from
over-smoothing~\cite{li2018deeper}. Our framework addresses both
limitations via Semantic Lifting (Section~3.2), which recovers
behavioral semantics for stripped nodes, and a Graphormer architecture
(Section~3.4), which captures global dependencies without the
over-smoothing bottleneck.

\subsection{LLM-Driven Code Understanding and Neuro-Symbolic Analysis}

Large Language Models have substantially expanded the analytical
capabilities available for code understanding. In the binary domain,
BinLLM~\cite{pan2025large} recovers variable names, function boundaries,
and high-level summaries from raw assembly, while
VulRepair~\cite{fu2022vulrepair} applies T5-based sequence-to-sequence
models to automated patch generation. Neuro-symbolic variants such as
GRACE~\cite{liu2023grace} convert CPGs into prompts under a
Retrieval-Augmented Generation loop, and
PDBERT~\cite{liu2024pre} pre-trains a Transformer encoder on
program dependency structures. These approaches share a common
limitation: they treat the LLM as either a black-box oracle or a
feature extractor, without any independent structural verification of
its outputs, leaving hallucinations unbounded in principle.

The gap widens in the industrial domain, where generic LLMs lack
knowledge of proprietary protocols and non-standard compiler artifacts, and
consequently generate elevated false-positive rates on multi-component
firmware. Our Reflexive Prompting pipeline (Section~3.2.4) addresses
this gap by introducing a structural oracle that rejects hallucinated data-flow edges before they enter the training corpus --- a verification contract absent from prior hybrid methods.

\subsection{Analysis of Opaque Industrial Firmware}

Ensuring transparency across the industrial supply chain is a 
prerequisite for defense-in-depth. Emerging regulatory
frameworks, including U.S.\ Executive Order
14028~\cite{eo14028} and the NIST SP~800-82
guidelines~\cite{nist_sp80082}, mandate rigorous software supply chain
transparency through mechanisms such as Software Bill of Materials
(SBOM) generation. The IEC~62443 standard~\cite{iec62443} further
regulates secure product development lifecycles in ICS environments.
In practice, industry tools such as Syft and Trivy map component
dependencies by parsing package manifests. This regulatory push creates 
the aforementioned Transparency Paradox: while standards demand deep 
visibility, industrial vendors frequently deliver opaque, closed-source 
firmware that lacks the package manifests (e.g., \texttt{package.json},
\texttt{conanfile.txt}) on which existing SCA tools depend.

On the analysis side, efforts to penetrate this opacity have encountered 
severe architectural bottlenecks.Static and symbolic execution platforms  angr~\cite{shoshitaishvili2016sok} face well-documented scalability 
limits when applied to monolithic firmware. 
While formal verification has been proposed for Programmable Logic Controller
(PLC) programs~\cite{zhang2024binary}, such approaches assume access 
to structured ladder-logic source code—an assumption that fails 
for stripped binaries. Conversely, dynamic analysis techniques, 
including firmware fuzzers~\cite{srivastava2019firmfuzz} and runtime monitors~\cite{tan2026operational}, hit the 'Environment Wall'
~\cite{muench2019dynamic}: emulating proprietary hardware peripherals 
for high-coverage dynamic testing remains prohibitively expensive and often impractical~\cite{wright2021challenges}.". Our SSCKG construction (Section~3.3) 
resolves this impasse. By operating entirely on static binary artifacts, 
it bypasses the hardware-emulation barrier of dynamic analysis, 
while simultaneously recovering the high-level behavioral semantics 
that traditional static tools discard.

\subsection{Code Representation and Software Knowledge Graphs}

Because transforming the CPG into the SSCKG is a primary
methodological contribution of this work (Section~3.3), it is
important to position it against existing code knowledge graphs.
CodeQL~\cite{youn2023declarative}, developed by GitHub, maps source-level ASTs
into a relational database and supports Datalog-style semantic
queries; it excels at pattern-based vulnerability detection for
open-source projects but operates exclusively on source code and
lacks abstract semantic or risk-propagation reasoning. Developer-centric
knowledge graphs such as DevKG link pull requests, commits, and
developer metadata to facilitate project-level analytics, yet they
encode no behavioral semantics of the code itself. More recent work
on code knowledge graphs has explored embedding function-level
summaries as node attributes, but these graphs remain anchored to
source-level identifiers and do not generalize to stripped
binaries.

In contrast, our SSCKG is an \emph{abstracted behavioral graph}
specifically designed for opaque firmware. It differs from the above
in three respects: (1)~entities are elevated from instruction-level
CPG nodes to behaviorally typed units via the Semantic Lifting stage,
enabling analysis of stripped binaries without source code;
(2)~relations encode multi-hop vulnerability propagation paths
(e.g., \texttt{taints}, \texttt{reaches}) drawn from a formal
eight-type ontology (Table~\ref{tab:relation_ontology}), rather than
raw syntactic edges; and (3)~the surjective transformation $\Phi$
(Section~3.3) provides a principled compression from millions of
instruction-level nodes to thousands of semantically meaningful
entities, making downstream Graphormer reasoning tractable.

\subsection{Summary and Positioning}

The preceding review reveals a consistent pattern: graph-based methods
offer structural rigor but lack semantic depth for stripped binaries;
LLM-based methods provide semantic insight but lack structural
grounding and domain specificity; industrial analysis tools face
hardware-emulation barriers; and existing code knowledge graphs do not
generalize beyond source-level representations.
Table~\ref{tab:related_comparison} synthesizes these observations across
five desiderata.

\begin{table}
  \caption{Comparison of representative approaches against five
    desiderata for industrial supply chain vulnerability detection.
    {\normalfont\checkmark}~= fully addressed;
    {\normalfont$\triangle$}~= partially addressed;
    {\normalfont$\times$}~= not addressed.}
  \label{tab:related_comparison}
  \centering
  \begin{tabular}{lccccc}
    \toprule
    Approach
      & \makecell{Structural\\Analysis}
      & \makecell{Semantic\\Understanding}
      & \makecell{Industrial\\Domain}
      & \makecell{Binary\\Ready}
      & \makecell{Formal\\Guarantee} \\
    \midrule
    VulDeePecker~\cite{li2018vuldeepecker}
      & $\times$     & $\times$     & $\times$     & $\times$     & $\times$ \\
    SySeVR~\cite{li2021sysevr}
      & \checkmark   & $\times$     & $\times$     & $\times$     & $\times$ \\
    Reveal~\cite{chakraborty2021deep}
      & \checkmark   & $\times$     & $\times$     & $\times$     & $\times$ \\
    DeepSeek-Coder~\cite{guo2024deepseek}
      & $\times$     & \checkmark   & $\times$     & $\triangle$  & $\times$ \\
    GRACE~\cite{liu2023grace}
      & \checkmark   & \checkmark   & $\times$     & $\times$     & $\times$ \\
    CodeQL~\cite{youn2023declarative}
      & \checkmark   & $\triangle$  & $\times$     & $\times$     & $\times$ \\
    angr~\cite{shoshitaishvili2016sok}
      & \checkmark   & $\times$     & $\triangle$  & \checkmark   & $\times$ \\
    \textbf{Ours}
      & \checkmark   & \checkmark   & \checkmark   & \checkmark   & \checkmark \\
    \bottomrule
  \end{tabular}
\end{table}

\noindent To the best of our knowledge, no existing work integrates
LLM-driven abstract interpretation with Graphormer-based graph
reasoning to construct a formally grounded, verifiable knowledge graph
for the opaque industrial software supply chain. The proposed framework
addresses this gap through four complementary mechanisms: a Galois
Connection that provides soundness guarantees for LLM-based semantic
recovery (Section~3.2), a Reflexive Prompting pipeline that eliminates
hallucinated abstractions (Section~3.2.4), a surjective SSCKG
transformation that encodes vulnerability propagation semantics
(Section~3.3), and a domain-adapted Graphormer that captures long-range
dependency chains beyond the reach of standard GNNs (Section~3.4).

\section{Methodology}

\subsection{Preliminaries and Problem Formulation}

The fundamental challenge in auditing the industrial software supply
chain is the absence of source code and debug symbols. To systematically
analyze these opaque binaries, our Supply Chain Analysis Agent (SCAA)
first parses the target executable (e.g., PE/ELF files or firmware
images) into a structural representation.

\begin{figure*}[t]
  \centering
  \includegraphics[width=\textwidth]{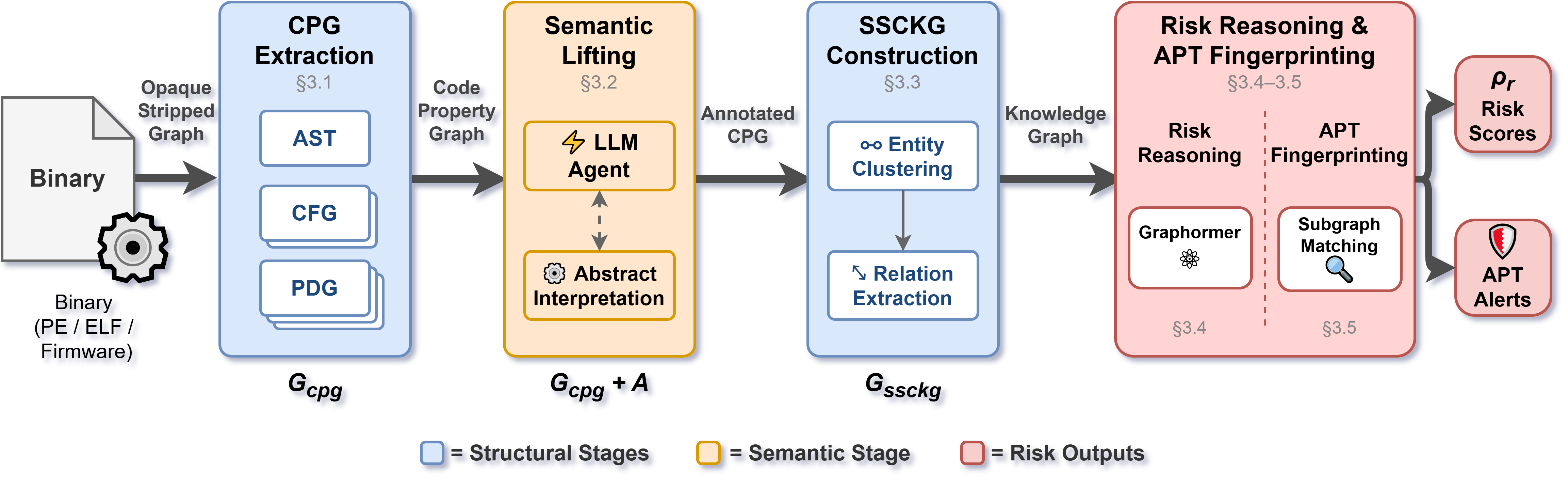}
  \caption{End-to-end pipeline of the proposed SCAA framework. A
    stripped binary is first parsed into a Code Property Graph
    ($G_{cpg}$, Section~3.1), then semantically lifted via a
    teacher--verifier--student loop (Section~3.2). The annotated CPG is
    transformed into a Software Supply Chain Knowledge Graph
    ($G_{ssckg}$, Section~3.3), on which a domain-adapted Graphormer
    performs risk reasoning and APT fingerprinting
    (Sections~3.4--3.5). Color coding: blue for structural stages,
    orange for semantic stages, red for risk outputs.}
  \label{fig:pipeline}
\end{figure*}

\textbf{Pipeline Overview.} To address the semantic gap inherent in
stripped binaries, we propose an end-to-end neuro-symbolic workflow
comprising four sequential stages (illustrated in
Figure~\ref{fig:pipeline}):

\begin{enumerate}
  \item \textbf{Binary $\rightarrow$ CPG Extraction} (Section~3.1):
    The executable is parsed into a Code Property Graph that unifies
    AST, CFG, and PDG representations.
  \item \textbf{Semantic Lifting} (Section~3.2): A fine-tuned LLM
    agent, governed by Abstract Interpretation theory, maps low-level
    instructions to a hierarchical domain of opaque industrial software security behaviors.
  \item \textbf{SSCKG Construction} (Section~3.3): The annotated CPG
    is transformed into a compact Knowledge Graph with typed
    vulnerability relations via entity clustering and relation
    extraction.
  \item \textbf{Risk Reasoning \& APT Fingerprinting}
    (Sections~3.4--3.5): A domain-adapted Graphormer captures
    long-range dependency chains and computes composite risk scores;
    a subgraph similarity algorithm detects zero-day threats by
    matching against known APT behavioral fingerprints.
\end{enumerate}

\textbf{Code Property Graph (CPG):} We adopt the standard CPG
formalism~\cite{yamaguchi2014modeling} as our foundational data structure
because it integrates Abstract Syntax Trees (AST), Control Flow
Graphs (CFG), and Program Dependence Graphs (PDG) into a single
queryable multigraph $G_{cpg}=(V,E)$ with edge set
$E = E_{ast}\cup E_{cfg}\cup E_{pdg}$, where nodes carry opcode and
type attributes and edges carry structural or data-flow labels ---
a unification essential for tracking cross-component vulnerabilities.

\textbf{The Semantic Gap Problem:} While $G_{cpg}$ mathematically
captures the topology of execution paths and data dependencies, it
inherently lacks high-level behavioral semantics when generated from
stripped binaries. A subgraph representing a proprietary industrial compiler 
operation or a hardware-controller state-machine translation remains semantically 
opaque; standard graph traversal algorithms cannot differentiate between
a legitimate network initialization and a malicious backdoor connection
based purely on structural features.

\textbf{Problem Formulation:} We formulate the task of Industrial Supply
Chain Vulnerability Detection as a two-stage neuro-symbolic learning
problem. Given a stripped binary artifact $\mathcal{B}$, our ultimate
objective is to learn a risk prediction function
$\mathcal{F}: \mathcal{B} \rightarrow [0,1]$. Because direct mapping is
intractable due to the aforementioned semantic gap, we decompose
$\mathcal{F}$ into two distinct operations:

\begin{enumerate}
  \item \textbf{Semantic Transformation ($\Phi$):} We must define a
    mapping function that elevates the structurally rigid but
    semantically sparse $G_{cpg}$ into a rich, queryable Software Supply
    Chain Knowledge Graph (SSCKG), representing
    ``behavior-dependency-risk'' relationships.
  \item \textbf{Risk Reasoning ($\Psi$):} We must define a graph
    reasoning function operating on the SSCKG to capture multi-hop,
    long-range dependencies and output a composite risk score $\rho_r$,
    enabling the detection of both known components and zero-day
    anomalous behaviors.
\end{enumerate}

\noindent Concretely, $\Phi$ is realized across Sections~3.2--3.3: the
Semantic Lifting stage (Section~3.2) maps concrete binary instructions
onto abstract behavioral labels in $\mathcal{A}$, while the SSCKG
Construction stage (Section~3.3) compresses those annotated instructions
into a compact, typed knowledge graph $G_\text{ssckg}$. The Risk
Reasoning function $\Psi$ is then realized in Sections~3.4--3.5, which
operate jointly on $G_\text{ssckg}$ to produce composite risk scores and
zero-day alert decisions. The following subsections detail each stage in
sequence. All non-standard symbols are introduced inline on first use.

\subsection{Semantic Lifting via Abstract Interpretation}

To bridge the semantic gap inherent in stripped binaries, we propose a
Semantic Lifting approach governed by the theory of Abstract
Interpretation. Unlike traditional pattern-matching techniques (e.g.,
YARA rules or byte-signature scanners) that fail on obfuscated or
previously unseen code, we model the semantic recovery process as a
principled mapping between the concrete execution space of the binary
and an abstract security domain. This formulation provides a
theoretical foundation for quantifying the fidelity of LLM-based code
understanding --- a guarantee absent from existing LLM-for-code
approaches that treat the model as a black-box oracle.

\subsubsection{Abstract Domain Definition}
Let $\mathcal{C}$ denote the concrete state space of the program
(registers and memory) and $\mathcal{A}$ denote the abstract domain
representing high-level security behaviors. We define $\mathcal{A}$ as a finite, domain-aware hierarchical lattice representing high-level security behaviors in opaque industrial software. To instantiate this lattice for our primary ICS case study, we derive the specific behavioral labels from the MITRE ATT\&CK for ICS matrix~\cite{mitre_attack_ics}; other domains would substitute domain-appropriate label sets while preserving the same three-tier structure. The hierarchy is organized as follows:

\begin{itemize}
  \item \textbf{Tier~1 (Macro-Behavior):} Five root categories ---
    \texttt{Network}, \texttt{Memory}, \texttt{Hardware},
    \texttt{FileSystem}, and \texttt{Cryptography}.
  \item \textbf{Tier~2 (Specific Action):} Fine-grained behavioral
    labels within each category. For example, \texttt{Network}
    decomposes into \texttt{Socket\_Init}, \texttt{Protocol\_Parse},
    \texttt{DNS\_Resolve}; \texttt{Hardware} includes
    \texttt{Register\_Read}, \texttt{Coil\_Write},
    \texttt{Firmware\_Update}.
  \item \textbf{Tier~3 (Risk Context):} Security-qualified variants
    that encode violation conditions, e.g.,
    \texttt{Unauthenticated\_Coil\_Write},
    \texttt{Unbounded\_Protocol\_Parse}.
\end{itemize}

\begin{figure}[t]
  \centering
  \includegraphics[width=\columnwidth]{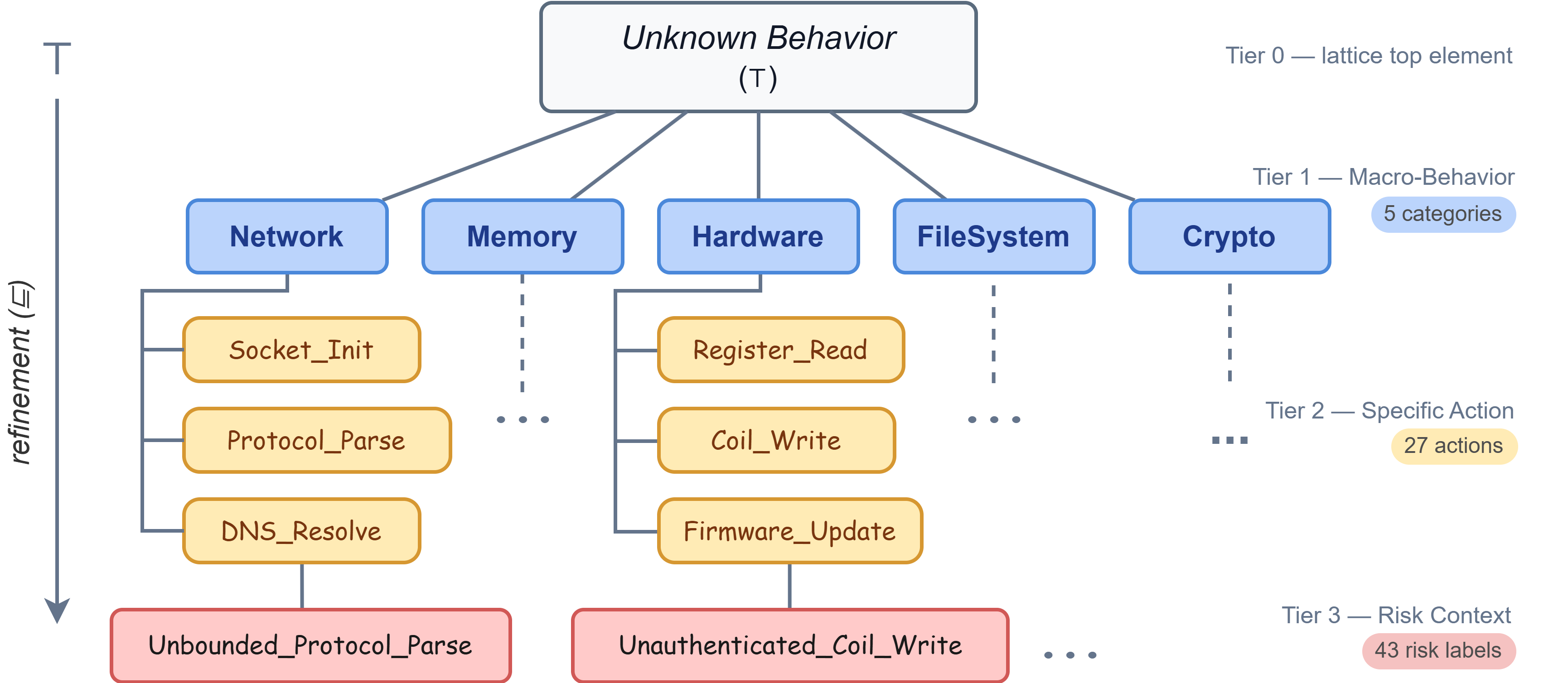}
  \caption{Three-tier abstract domain $\mathcal{A}$ derived from MITRE
    ATT\&CK for ICS. Tier~1 comprises 5 macro-behavior categories;
    Tier~2 refines these into 27 specific actions; Tier~3 qualifies 43
    risk-contextualized labels. The partial order $\sqsubseteq$ follows
    the containment hierarchy, with $\top$ (Unknown Behavior) as the
    top element. Only representative branches are shown for clarity.}
  \label{fig:abstract_domain}
\end{figure}

\noindent In total, $\mathcal{A}$ comprises 5 macro-categories, 27
specific actions, and 43 risk-contextualized labels. The partial order
$\sqsubseteq$ on $\mathcal{A}$ follows the containment hierarchy: a
Tier~3 label is below its parent Tier~2 action, which is below its
Tier~1 category, with $\top$ denoting the universal ``Unknown
Behavior'' element. This lattice structure ensures that
over-approximation (mapping to a coarser category) preserves soundness,
while under-approximation (missing a risk context) constitutes a
violation. Figure~\ref{fig:abstract_domain} depicts representative
branches of this three-tier lattice, illustrating how concrete binary
behaviors are progressively abstracted into security-relevant categories.

\subsubsection{Galois Connection Formulation}
We establish a Galois Connection $(\alpha, \gamma)$~\cite{cousot1977abstract} between the
concrete and abstract spaces:

\begin{equation}
  \alpha : \mathcal{P}(\mathcal{C}) \rightarrow \mathcal{A},
  \quad
  \gamma : \mathcal{A} \rightarrow \mathcal{P}(\mathcal{C})
\end{equation}

\noindent where $\alpha$ is the abstraction function and $\gamma$ is the
concretization function. The pair $(\alpha, \gamma)$ constitutes a valid
Galois Connection if and only if $\alpha(c) \sqsubseteq a \iff c \in
\gamma(a)$ for all $c \in \mathcal{P}(\mathcal{C})$ and $a \in
\mathcal{A}$~\cite{cousot1977abstract}. Because the concrete state space
$\mathcal{C}$ (registers and memory) is unbounded, $\gamma(a)$ is not
computationally tractable. Compounding this, the concrete
transition function $F$ (the binary execution) is opaque for stripped
binaries, making $\alpha \circ F$ inaccessible. We therefore employ a
Large Language Model (LLM) agent, specifically a
fine-tuned Qwen3-7B~\cite{yang2025qwen3} (via QLoRA~\cite{dettmers2023qlora}), to approximate the abstract transition
function $F^\sharp$. The objective is to ensure that the LLM's
interpretation satisfies the soundness condition:

\begin{equation}
  \alpha(F(c)) \sqsubseteq F^\sharp(\alpha(c))
\end{equation}

\noindent where $\sqsubseteq$ represents the partial order in the risk
hierarchy defined above. Intuitively, this condition requires that the
LLM's behavioral summary must \emph{cover} the true concrete behavior
--- it may over-approximate (e.g., labeling a benign function as
\texttt{Network} when only \texttt{Socket\_Init} applies) but must not
\emph{miss} security-relevant behaviors.

\subsubsection{Empirical Soundness Verification}
Due to the probabilistic nature of LLMs, absolute mathematical
soundness is intractable. We instead define the \emph{Empirical
Violation Rate} (EVR, formally stated in Eq.~10, Section~4.2.3) as the
fraction of curated Golden Set functions whose LLM-predicted abstract
label fails to cover the expert ground truth in the lattice
$\mathcal{A}$. A lower EVR indicates that the approximation
$F^\sharp$ is closer to the ideal abstraction function. The
full empirical validation --- including per-category breakdown, error
taxonomy, and comparison against baseline LLMs --- is reported in
Section~4.4 (RQ4).

\subsubsection{Reflexive Prompting: A Teacher--Student Verification Loop}
To minimize the EVR and enforce structural consistency, the SCAA
employs a Reflexive Prompting mechanism comprising a teacher--student
pipeline with graph-based verification. Unlike standard knowledge
distillation, this loop incorporates a structural oracle (the Joern
CPG server~\cite{yamaguchi2014modeling}) as an independent verifier, ensuring that the LLM's
semantic summaries remain grounded in the program's actual data-flow
topology.

The procedure operates as follows:

\begin{enumerate}
  \item \textbf{Teacher Generation:} A large-capacity teacher model
    (DeepSeek-v3) receives a decompiled function along with its CPG
    context (control-flow predecessors, callee signatures) and produces
    an initial semantic summary, mapping each basic block to a label in
    $\mathcal{A}$. The prompt template follows a structured format:
    \texttt{[Context: Control Flow Predecessors]},
    \texttt{[Instruction: Map to Domain~$\mathcal{A}$]},
    \texttt{[Constraint: Output strict JSON]}.
  \item \textbf{Structural Verification:} The Joern server executes a
    graph traversal $T$ based on the teacher's claimed data-flow paths.
    If the traversal returns \texttt{UNSAT} (i.e., the teacher
    hallucinated a data-flow edge that does not exist in the CPG), a
    negative correction signal is generated, and the teacher's summary
    is rejected for that block.
  \item \textbf{Student Fine-Tuning:} The structurally verified
    summaries form a training corpus of 15,000 instruction--behavior
    pairs. We fine-tune the local Qwen3-7B agent via QLoRA (rank~16,
    $\alpha$=32) with a learning rate of $2 \times 10^{-4}$. Training
    was halted after 5 epochs when validation loss plateaued at 0.12.
\end{enumerate}

\begin{figure}[t]
  \centering
  \includegraphics[width=\columnwidth]{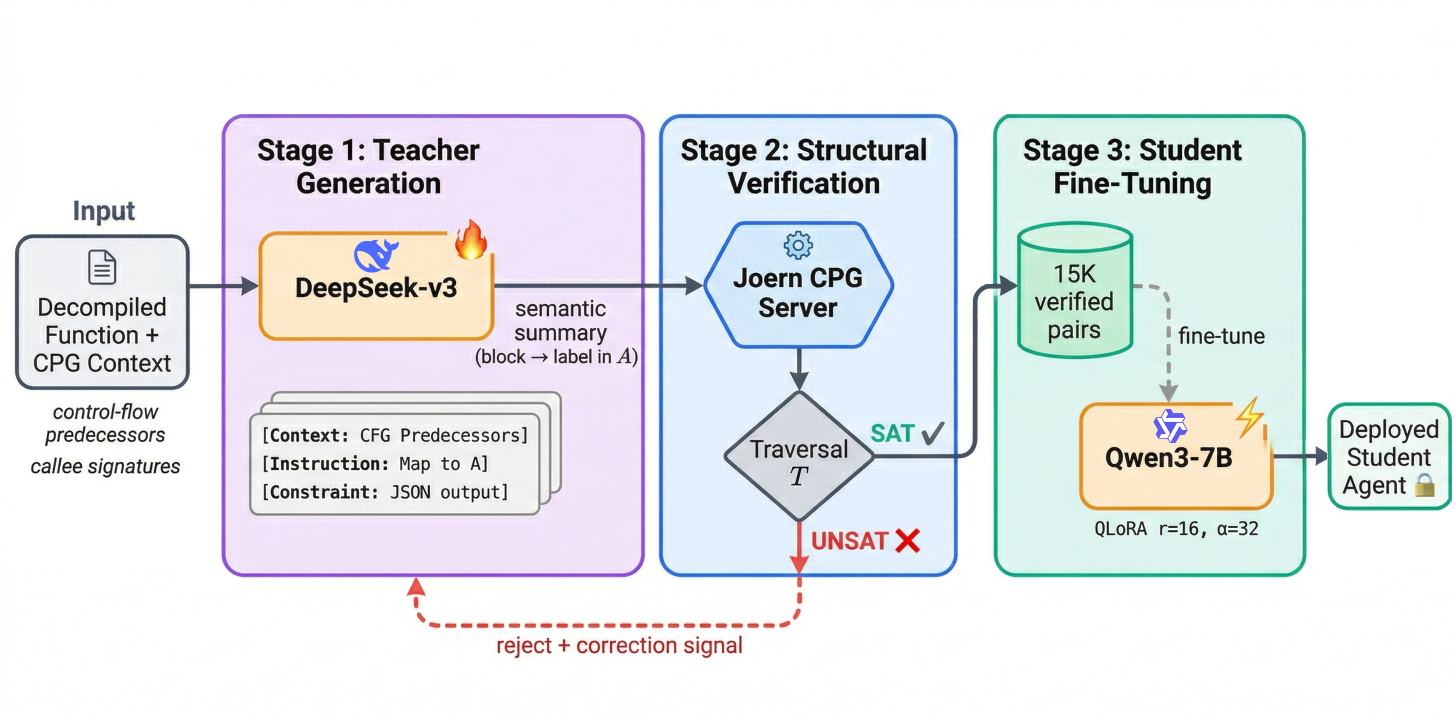}
  \caption{Reflexive Prompting pipeline. The teacher model
    (DeepSeek-v3) generates semantic summaries, which are verified
    against the CPG by the Joern structural oracle. Summaries that fail
    the SAT check (15.8\% rejection rate) are discarded. The 15,000
    verified instruction--behavior pairs are used to fine-tune the
    student model (Qwen3-7B) via QLoRA (rank=16, $\alpha$=32,
    lr=$2 \times 10^{-4}$, 5 epochs).}
  \label{fig:reflexive_prompting}
\end{figure}

\noindent This three-stage pipeline ensures that the student model
inherits the teacher's broad semantic understanding while the
structural verifier eliminates hallucinated abstractions that violate
the CPG's ground-truth topology. We select Qwen3-7B as the student
model for its strong performance on code understanding benchmarks
while remaining deployable on a single GPU (24~GB VRAM) --- a
practical requirement for offline industrial environments where cloud
API access is restricted. Figure~\ref{fig:reflexive_prompting}
summarizes the complete teacher--verifier--student loop, including the
rejection rate and key training hyperparameters.

\begin{algorithm}[t]
\caption{Reflexive Prompting Corpus Construction}
\label{alg:reflexive_prompting}
\DontPrintSemicolon
\KwIn{Function set $\mathcal{F} = \{f_i\}$; CPG oracle \textsc{Joern};
      teacher model $M_T$ (DeepSeek-v3); student model $M_S$ (Qwen3-7B)}
\KwOut{Fine-tuned student model $M_S^*$}
$\mathcal{D} \leftarrow \emptyset$\;
\ForEach{$f_i \in \mathcal{F}$}{
  $\ell_T \leftarrow M_T\bigl(f_i,\; \textsc{CPG\_Context}(f_i)\bigr)$
  \tcp*{Teacher: generate semantic summary}
  $r \leftarrow \textsc{Joern.Traverse}\bigl(f_i,\;
      \ell_T.\textit{data\_flow\_claims}\bigr)$
  \tcp*{Verifier: check structural consistency}
  \If{$r \neq \texttt{UNSAT}$}{
    $\mathcal{D} \leftarrow \mathcal{D} \cup \{(f_i,\, \ell_T)\}$
    \tcp*{Accept verified pair ($\approx$84.2\% of inputs)}
  }
  \Else{
    \textbf{discard} $\ell_T$
    \tcp*{Reject hallucinated data-flow edges (15.8\%)}
  }
}
$M_S^* \leftarrow \textsc{QLoRA\_Finetune}(M_S,\; \mathcal{D},\;
    \text{rank}{=}16,\; \alpha{=}32,\; \text{lr}{=}2{\times}10^{-4},\;
    \text{epochs}{=}5)$\;
\Return $M_S^*$
\end{algorithm}

\noindent The verified $\langle f_i, \ell_T \rangle$ pairs produced by
the fine-tuned student model $M_S^*$ constitute the behavioral
annotations that seed both downstream stages: the abstract labels in
$\mathcal{A}$ drive entity clustering in Section~3.3, and the
behavioral summaries $\text{desc}(v)$ feed the inherent risk computation
$\rho_\text{inherent}(v)$ in Section~3.4 (Eq.~8).

\subsection{SSCKG Construction: The Semantic Graph Transformation}

With each basic block now annotated with behavioral labels from
$\mathcal{A}$, the next stage transforms the fine-grained CPG into a
compact, queryable Software Supply Chain Knowledge Graph (SSCKG). This
transformation serves two purposes: it reduces the graph from millions
of instruction-level nodes to thousands of semantically meaningful
entities, and it introduces typed relations that encode vulnerability
propagation paths not explicit in the original CPG. We formalize this
as a surjective graph transformation function $\Phi$:

\begin{equation}
  \Phi : G_{cpg}(V, E) \rightarrow G_{ssckg}(\mathcal{E}, \mathcal{R})
\end{equation}

\noindent We note that $\Phi$ is \emph{not} an isomorphism: it
deliberately introduces semantic information (behavioral labels, risk
relations) that is absent in the source CPG, while collapsing
structurally redundant nodes. The transformation consists of two
concurrent processes.

\subsubsection{Entity Elevation ($\Phi_{\mathcal{E}}$)}
Nodes $v \in V$ in the CPG are clustered and mapped to high-level
entities $e \in \mathcal{E}$ via a hybrid algorithm that combines
structural boundaries with semantic similarity:

\begin{enumerate}
  \item \textbf{Structural Collapse (Rule-Based):} All AST nodes
    belonging to a single basic block or a single function body are
    collapsed into one entity $e_i$. This step preserves program
    structure boundaries and reduces the node count by approximately
    two orders of magnitude, following standard program slicing
    conventions~\cite{korel1998dynamic}.
  \item \textbf{Semantic Clustering (Embedding-Based):} For external
    library calls and cross-module functions, we generate
    Sentence-BERT~\cite{reimers2019sentence} embeddings of the
    behavioral summaries produced by the Semantic Lifting stage
    (Section~3.2). We then apply DBSCAN~\cite{ester1996density}
     ($\varepsilon$=0.3,\\ \texttt{min\_samples}=2) to cluster 
    semantically equivalent but syntactically distinct functions. 
    For example, \texttt{s7\_read\_req} and \texttt{read\_modbus\_register} ---
    despite having different implementations --- both cluster into the
    entity \texttt{PLC\_Read\_Request} (the ICS-domain label for this 
    behavioral class; analogous entity names would apply in automotive 
    or aerospace instantiations), because their behavioral
    summaries share the Tier~2 label \texttt{Register\_Read}. We
    select DBSCAN over $k$-means because the number of semantic
    clusters is not known a priori and DBSCAN naturally identifies
    noise points (functions with unique behaviors).
\end{enumerate}

\subsubsection{Relation Extraction ($\Phi_{\mathcal{R}}$)}
Edges in the CPG are transformed into semantically typed relations.
We define a relation ontology $\Sigma_{\mathcal{R}}$ comprising eight
relation types organized into three categories, summarized in
Table~\ref{tab:relation_ontology}.

\begin{table}
  \caption{SSCKG Relation Ontology ($\Sigma_{\mathcal{R}}$). Eight
    typed relations in three categories, with their derivation source
    from the CPG edge sets.}
  \label{tab:relation_ontology}
  \begin{tabular}{llll}
    \toprule
    Category & Relation Type & Source & Semantics \\
    \midrule
    \multirow{3}{*}{Structural}
      & \texttt{calls}          & $E_{ast}$        & Function invocation \\
      & \texttt{depends\_on}    & $E_{ast}$        & Build/link dependency \\
      & \texttt{imports}        & $E_{ast}$        & Library inclusion \\
    \midrule
    \multirow{2}{*}{Data Flow}
      & \texttt{reads\_from}    & $E_{pdg}$        & Memory/register read \\
      & \texttt{writes\_to}     & $E_{pdg}$        & Memory/register write \\
    \midrule
    \multirow{3}{*}{Vulnerability}
      & \texttt{taints}         & $E_{pdg}$ + NVD  & Tainted data propagation \\
      & \texttt{reaches}        & $E_{pdg}$ + NVD  & Reachability to sensitive op \\
      & \texttt{vulnerable\_to} & NVD match        & Entity $\rightarrow$ CVE mapping \\
    \bottomrule
  \end{tabular}
\end{table}

\noindent The eight-relation ontology was derived by cross-mapping the
tactic categories of MITRE ATT\&CK
against the CWE vulnerability taxonomy from the NVD. Each of the three
structural relations corresponds to a CPG edge type; each data-flow
relation encodes a program dependence direction; and the three
vulnerability relations cover the full spectrum from upstream taint
propagation to direct CVE matching. Relations outside the scope of 
device-level opaque software execution --- such as multi-node network 
topology links or high-level session states --- are intentionally 
excluded and constitute a known limitation acknowledged in Section~5.2.

\noindent Relation types are assigned deterministically: structural
relations are inherited directly from the CPG edge types
($E_{ast}$, $E_{cfg}$); data-flow relations are derived from
$E_{pdg}$ edges; and vulnerability relations are inferred by
combining the Semantic Lifting labels with known vulnerability
patterns from the NVD. For instance, a data dependency edge
$e_{pdg}$ connecting a user input node to a sensitive API node is
transformed into a risk relation triplet:
\begin{equation}
  r_{flow} = \langle \Phi_{\mathcal{E}}(v_{input}),\;
    \texttt{taints},\;
    \Phi_{\mathcal{E}}(v_{sink}) \rangle \in \mathcal{R}
\end{equation}

\begin{figure*}[t]
  \centering
  \includegraphics[width=\textwidth]{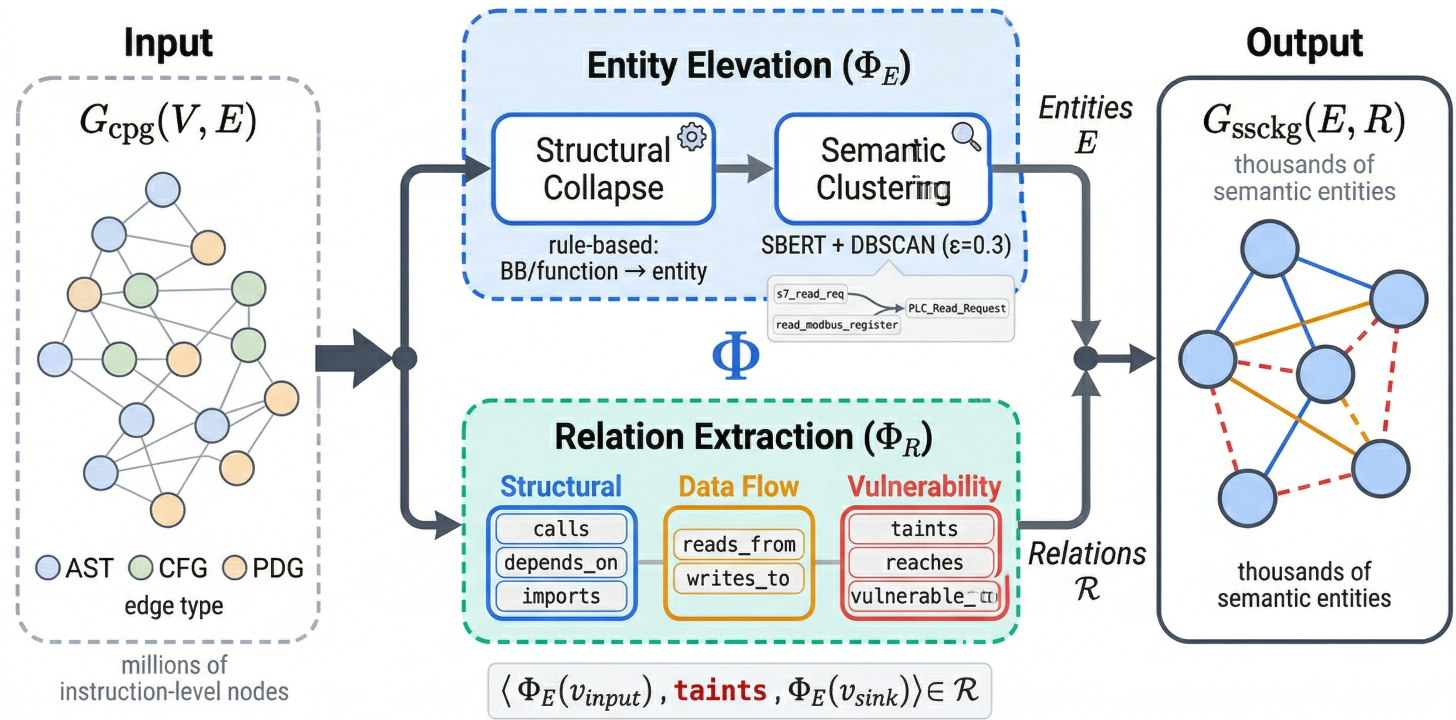}
  \caption{Concrete example of the surjective transformation
    $\Phi: G_{cpg} \rightarrow G_{ssckg}$. \textbf{Left:} A CPG
    fragment ($\sim$20 nodes) from a Modbus protocol handler, with
    AST (solid), CFG (dashed), and PDG (dotted) edges.
    \textbf{Right:} The resulting SSCKG fragment ($\sim$5 entities)
    after Entity Elevation and Relation Extraction, showing typed
    relations from $\Sigma_{\mathcal{R}}$ (Table~\ref{tab:relation_ontology}).
    Arrows indicate which CPG nodes collapse into each SSCKG entity;
    for instance, \texttt{s7\_read\_req} and
    \texttt{read\_modbus\_register} cluster into
    \texttt{PLC\_Read\_Request}.}
  \label{fig:ssckg_example}
\end{figure*}

\noindent This typed relation ontology distinguishes our SSCKG from
generic code knowledge graphs (e.g., those constructed by
CodeQL~\cite{youn2023declarative}), which lack explicit vulnerability propagation
semantics. To make this transformation concrete,
Figure~\ref{fig:ssckg_example} presents a before-and-after example
showing how a CPG fragment is compressed and enriched into an SSCKG
subgraph with typed vulnerability relations.

\noindent The resulting $G_\text{ssckg}(\mathcal{E}, \mathcal{R})$ is
the sole input consumed by both subsequent stages: Section~3.4 uses its
entity set $\mathcal{E}$ and typed relations $\mathcal{R}$ to drive
Graphormer attention and composite risk scoring, while Section~3.5
matches subgraphs of $G_\text{ssckg}$ against APT fingerprints in
embedding space.

\subsection{Graph Representation Learning via Graphormer}

With the SSCKG constructed, the next stage identifies latent risks by
capturing complex, non-local dependencies between entities. As
discussed in Section~2.1, message-passing GNNs suffer from $k$-hop
receptive fields and over-smoothing~\cite{li2018deeper} when
propagation chains span 10--15 hops, as is common in industrial
supply chains. To overcome this limitation, we employ a Graphormer
architecture~\cite{ying2021transformers}. Unlike message-passing GNNs,
Graphormer utilizes a Transformer-based global attention mechanism,
allowing every node to attend to every other node in a single layer,
regardless of topological distance. This architectural choice is
motivated by the observation that vulnerability propagation in supply
chains is fundamentally a \emph{global} property: a compromised
upstream library affects all downstream consumers, irrespective of
their position in the call graph.

The node update rule for the $l$-th layer follows the standard
Graphormer formulation:

\begin{equation}
  \mathbf{z}_v^{(l+1)} = \text{LayerNorm}\!\left(
    \mathbf{z}_v^{(l)} +
    \text{FFN}\!\left(\text{MultiHeadAttn}(\mathbf{z}^{(l)})\right)
  \right)
\end{equation}

\subsubsection{Domain-Specific Edge Encoding}
While Eq.~4 is the standard Graphormer update, our key adaptation lies
in the attention bias computation. In the original
Graphormer~\cite{ying2021transformers}, the attention score $A_{ij}$
is biased solely by the shortest-path distance $\phi(v_i, v_j)$.
We extend this by modulating the bias with the \emph{semantic weight}
$w_{vu}$ of the edge, derived from the relation ontology
$\Sigma_{\mathcal{R}}$ defined in Section~3.3:

\begin{equation}
  A_{ij} = \frac{(\mathbf{z}_i \mathbf{W}_Q)(\mathbf{z}_j \mathbf{W}_K)^\top}{\sqrt{d}}
    + b_{\phi(v_i, v_j)} + b_{\varphi(e_{ij})} \cdot w_{vu}
\end{equation}

\begin{figure}[t]
  \centering
  \includegraphics[width=\columnwidth]{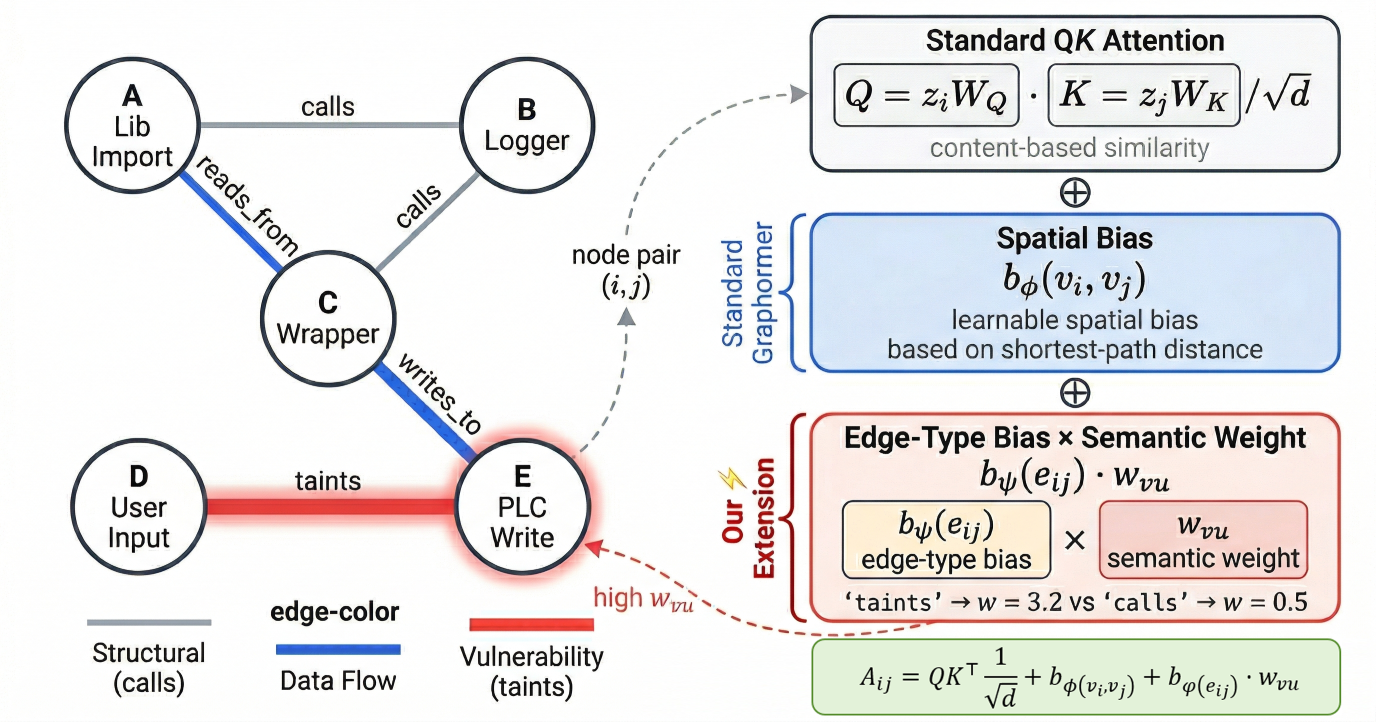}
  \caption{Domain-specific attention bias in a single Graphormer
    head. \textbf{Top:} Standard Graphormer biases attention by
    shortest-path distance $b_{\phi}$ only. \textbf{Bottom:} Our
    extension adds an edge-type bias $b_{\varphi}$ modulated by the
    semantic weight $w_{vu}$ from $\Sigma_{\mathcal{R}}$, amplifying
    vulnerability-relevant paths (\texttt{taints}, \texttt{reaches})
    over structural edges (\texttt{calls}, \texttt{imports}). Edge
    colors encode relation type; line thickness reflects attention
    magnitude.}
  \label{fig:graphormer_arch}
\end{figure}

\noindent Figure~\ref{fig:graphormer_arch} contrasts the standard
Graphormer attention bias (spatial only) with our extended formulation,
illustrating how edge-type and semantic weight jointly modulate the
attention scores on a representative SSCKG subgraph.

\noindent Here, $b_{\phi}$ is the learnable spatial bias and
$b_{\varphi}$ is the learnable edge-type bias. The semantic weight
$w_{vu}$ is assigned based on relation type: a \texttt{taints} edge
connecting user input to a PLC hardware interrupt receives a
substantially higher attention bias than a \texttt{calls} edge between
two benign logging functions. This domain-specific weighting ensures
that the model prioritizes vulnerability-relevant paths during
attention computation, rather than treating all graph edges uniformly.

\subsubsection{Composite Risk Score}
Based on the learned node embeddings $\mathbf{z}_v$, we compute the
Composite Risk Score $\rho_r(v)$ for each entity. This score
aggregates both the inherent vulnerability of the node and the
contextual risk propagated from its neighbors:

\begin{equation}
  \rho_r(v) = \beta \cdot \rho_{\text{inherent}}(v)
    + (1 - \beta) \cdot \frac{1}{|\mathcal{N}(v)|}
      \sum_{u \in \mathcal{N}(v)} w_{vu} \cdot \rho_r(u)
\end{equation}

\noindent Eq.~7 is solved via power iteration: beginning from
$\rho_r^{(0)}(v) = \rho_{\text{inherent}}(v)$, each step updates
all entity scores simultaneously until $\|\rho_r^{(t+1)} -
\rho_r^{(t)}\|_1 < 10^{-6}$ (typically 20--30 iterations).
Convergence is guaranteed by the Perron--Frobenius theorem provided
the weight matrix $W = \{w_{vu}\}$ is row-stochastic after
per-node normalization, which our relation-weight assignment satisfies
by construction.
We further note that Eq.~7 structurally resembles Personalized
PageRank~\cite{bianchini2005inside} with $\rho_{\text{inherent}}$
serving as the personalization vector. This connection is deliberate:
PageRank's iterative diffusion naturally models how risk propagates
through dependency chains, while the learned attention weights $w_{vu}$
from the Graphormer replace the uniform transition probabilities with
semantically informed edge strengths.

The inherent risk $\rho_{\text{inherent}}(v)$ is computed via
Sentence-BERT~\cite{reimers2019sentence} semantic alignment between
the entity's behavioral summary (from Section~3.2) and known CVE
descriptions from the NVD (the same Sentence-BERT encoder used for
behavioral clustering in Section~3.3.1; embeddings are computed once
during SSCKG construction and reused here without recomputation).
Specifically:

\begin{equation}
  \rho_{\text{inherent}}(v) = \max_{c \in \text{CVE}_{\text{ICS}}}
    \cos\!\left(\text{SBERT}(\text{desc}(v)),\;
    \text{SBERT}(\text{desc}(c))\right)
\end{equation}

\noindent where $\text{desc}(v)$ is the Semantic Lifting summary of
entity $v$ and $\text{desc}(c)$ is the textual description of CVE
entry $c$. The damping factor $\beta$ is set to 0.15, following the
standard PageRank convention, and validated via grid search on the
development set (Section~4).

\noindent The Graphormer node embeddings $\mathbf{z}_v$ produced by
this stage serve a dual role: the composite risk scores $\rho_r(v)$
drive the vulnerability detection decisions reported in Section~4, while
the embedding vectors themselves are passed directly to the APT
fingerprinting module in Section~3.5, where cosine similarity in
embedding space replaces intractable structural subgraph isomorphism.

\subsection{Zero-Day Threat Detection via APT Fingerprinting}

The preceding stages address known (1-day) vulnerability detection.
However, a critical requirement for industrial security is the detection
of Advanced Persistent Threats (APTs) and zero-day exploits that utilize
``legitimate but abnormal'' logic --- for instance, a specific sequence
of PLC register writes combined with a covert network socket opening ---
that evades standard signature-based detection.

We formulate this as a Subgraph Similarity Search problem operating in
the Graphormer embedding space, rather than on raw graph structure. This
design choice is motivated by a key observation: while structural
subgraph isomorphism is NP-hard in
general~\cite{cordella2004sub}, computing cosine similarity between
fixed-dimensional embeddings is $O(|\mathcal{V}_{\text{apt}}| \cdot
|\mathcal{V}_{\text{target}}|)$, which is tractable for the entity-level
SSCKG (typically $10^3$--$10^4$ nodes after Entity Elevation).

\subsubsection{APT Fingerprint Construction}
We maintain a repository of APT Fingerprints, represented as abstract
subgraphs $\mathcal{G}_{\text{apt}}$, constructed via a semi-automated
extraction pipeline:

\begin{enumerate}
  \item \textbf{Source Data:} To model advanced threats targeting opaque 
    industrial software, we utilize the MITRE Engenuity ICS APT
    evaluation dataset. This provides documented behavioral traces for 
    sophisticated supply chain and logic-level attacks (e.g., Stuxnet, 
    Triton, and HAVEX).~\cite{mitre_engenuity_ics}.
  \item \textbf{SSCKG Generation:} Known malware binaries are processed
    through the full SCAA pipeline (Sections~3.2--3.4) to generate
    their SSCKGs, producing entity-level behavioral graphs with typed
    relations.
  \item \textbf{Expert Pruning:} Since malware SSCKGs contain
    substantial boilerplate code (e.g., runtime initialization, benign
    library calls), domain experts manually pruned each graph to isolate
    the core malicious subgraphs $\mathcal{G}_{\text{apt}}$. For
    example, the Stuxnet fingerprint retains only the subgraph where a
    malicious DLL injection leads to rogue \texttt{s7comm} packets
    targeting specific PLC models, discarding unrelated system
    initialization routines.
\end{enumerate}

\subsubsection{Similarity Metric}
To detect a potential zero-day threat, we compute the semantic
similarity between the target software's SSCKG
$\mathcal{G}_{\text{target}}$ and each fingerprint in the repository.
The metric is defined as the average best-match cosine similarity:

\begin{equation}
  \text{Sim}(\mathcal{G}_{\text{target}}, \mathcal{G}_{\text{apt}}) =
    \frac{1}{|\mathcal{V}_{\text{apt}}|}
    \sum_{v \in \mathcal{V}_{\text{apt}}}
      \max_{u \in \mathcal{V}_{\text{target}}}
        \cos(\mathbf{z}_u, \mathbf{z}_v)
\end{equation}

\noindent where $\mathbf{z}_u$ and $\mathbf{z}_v$ are the Graphormer
embeddings from Section~3.4. We adopt this asymmetric formulation
(averaging over the fingerprint, maximizing over the target) because an
APT fingerprint captures a \emph{necessary} behavioral pattern: every
node in the fingerprint must find a semantic match in the target, but
the target may contain additional benign functionality. This design is
robust against code obfuscation because it operates on semantic
embeddings rather than raw structural features, allowing the system to
identify functionally equivalent attack patterns even when the
underlying binary implementation varies.

\subsubsection{Threshold Determination}
The alert threshold $\tau_{\text{apt}}$ is not a manually tuned
constant. We determine $\tau_{\text{apt}}$ empirically via grid search
on the validation splits of the Big-Vul and InduVul datasets,
optimizing Youden's J statistic ($J = \text{TPR} - \text{FPR}$) 
\cite{youden1950index} on the ROC curve. This criterion maximizes 
the True Positive Rate (recall for zero-day threats) 
while strictly constraining the False Positive Rate
to $\leq 5\%$, a requirement driven by the operational reality that
high FPR causes ``alert fatigue'' in industrial
settings~\cite{tariq2025alert}. The resulting threshold is
global across all fingerprint classes; per-fingerprint thresholds did
not yield statistically significant improvements in our experiments
(Section~4). If $\text{Sim} > \tau_{\text{apt}}$, the system triggers
a high-severity alert for manual forensic review.

\medskip
\noindent\textbf{End-to-End Information Flow.}
To make the stage interdependencies explicit, Figure~\ref{fig:pipeline}
traces a single stripped binary $\mathcal{B}$ through all five
transformations. First, $\mathcal{B}$ is parsed into a Code Property
Graph $G_{cpg}$ (Section~3.1). Second, each function in $G_{cpg}$ is
annotated with a behavioral label from the abstract security lattice
$\mathcal{A}$, and structurally verified instruction--behavior pairs are
used to fine-tune the student lifting agent $M_S^*$ (Section~3.2).
Third, $M_S^*$-annotated nodes are elevated into SSCKG entities and
connected via typed vulnerability relations, yielding $G_\text{ssckg}$
(Section~3.3). Fourth, a domain-adapted Graphormer operates on
$G_\text{ssckg}$ to produce per-entity embeddings $\mathbf{z}_v$ and
composite risk scores $\rho_r(v)$ (Section~3.4). Fifth, $\mathbf{z}_v$
vectors are compared against pre-built APT fingerprints via Eq.~9 to
detect zero-day threat patterns (Section~3.5). Each stage's output is
precisely the input consumed by the next; no stage requires information
that has not been produced by a prior step.

\section{Experimental Evaluation}

\subsection{Research Questions (RQs)}

To rigorously evaluate the proposed framework, we structure our
experiments around six research questions. We first establish baseline
superiority (RQ1--RQ2), then evaluate novel capabilities (RQ3--RQ4),
assess robustness (RQ5), and finally validate deployment readiness on
real industrial infrastructure (RQ6).

\begin{itemize}
  \item \textbf{RQ1 (Comparative Efficacy):} Does the proposed
    SSCKG-Graphormer framework outperform state-of-the-art (SOTA) static
    analysis and LLM-based methods in detecting vulnerabilities within
    opaque industrial software?
  \item \textbf{RQ2 (Ablation Study):} How much do the Semantic Lifting
    (LLM-based abstraction) and Graphormer (Global Attention) components
    individually contribute to detection performance?
  \item \textbf{RQ3 (Zero-Day \& APT Detection):} Can the system
    identify ``unknown'' threats (0-day vulnerabilities) and APT
    fingerprints that deviate from standard NVD patterns?
  \item \textbf{RQ4 (Semantic Lifting Fidelity):} How accurately does
    the fine-tuned LLM agent recover high-level security behaviors from
    stripped binaries, and where does it fail?
  \item \textbf{RQ5 (Sensitivity Analysis):} How sensitive is the
    framework to its key hyperparameters ($\beta$, $\tau_{\text{apt}}$,
    DBSCAN $\varepsilon$)?
  \item \textbf{RQ6 (Industrial Practicability and Real-World Efficacy):}
    How does the framework perform in a real-world case study of opaque 
    industrial environments (measured via throughput, False Positive Rate 
    reduction, and detection coverage of known high-impact CVEs on a hybrid
    virtual--physical testbed)?
\end{itemize}

\subsection{Experimental Setup}

\subsubsection{Datasets}
We evaluate on three datasets of increasing opacity and domain complexity.
Table~\ref{tab:datasets} summarizes their key statistics.

\begin{table}[h]
  \caption{Dataset Statistics}
  \label{tab:datasets}
  \begin{tabular}{lrrrl}
    \toprule
    Dataset & Total & Vuln. & Imbalance & Source \\
    \midrule
    Big-Vul           & 188,636 & 10,900 & 1:16  & Open-source C/C++ \\
    NVD-Precise (Ind.)& 3,218   & 487    & 1:5.6 & ICS CVEs \\
    InduVul-Dataset   & 1,247   & 59     & 1:20  & Real ICS firmware \\
    \bottomrule
  \end{tabular}
\end{table}

\begin{itemize}
  \item \textbf{Big-Vul}~\cite{fan2020ac}: A large-scale C/C++
    vulnerability dataset (188,636 functions, 10,900 vulnerable)
    collected from 348 open-source GitHub projects, transparent, 
    source-level software. We use Big-Vul to pre-training the Graphormer 
    on generalized vulnerability propagation and to establish a baseline 
    demonstrating how severely traditional methods degrade when transitioning 
    from transparent code to opaque binaries.
  \item \textbf{NVD-Precise (Industrial)}: A curated subset of 3,218
    functions extracted from ICS-related CVEs published by Siemens,
    Schneider Electric, and Rockwell Automation (advisories 2018--2025).
    Source code is compiled with three representative industrial
    toolchains (Tasking v6.3, Keil MDK-ARM v5.38, GCC-ARM 12.2) at
    optimization level \texttt{-O2}, then stripped of all debug symbols
    and section names to simulate real-world deployment conditions. This
    dataset serves as the primary training and validation source for the
    Semantic Lifting agent.
  \item \textbf{InduVul-Dataset}:\footnote{Dataset and testbed
    configuration publicly available at
    \url{https://github.com/Mewtwoz/InduVul-Dataset-testbed}.}
    Our constructed benchmark comprising
    1,247 stripped firmware functions extracted from 50 real-world ICS
    firmware images (Siemens S7-1200/1500, Rockwell CompactLogix,
    Schneider M340). Functions are labeled through a combination of
    known CVE mapping and manual expert review. The class imbalance
    ratio is approximately 1:20 (vulnerable to benign), reflecting the
    natural distribution in production firmware. This dataset serves as
    the held-out test benchmark for all RQs except where noted.
\end{itemize}

\noindent All datasets are split 70/10/20 (train/validation/test) with
stratified sampling to preserve class ratios within each split. No
function from the same binary appears in both training and test splits,
preventing data leakage through shared compilation artifacts.

\subsubsection{Baselines}
We compare against five representative methods spanning four
modalities: \textbf{SySeVR}~\cite{li2021sysevr} (slice + BiLSTM),
\textbf{Reveal}~\cite{chakraborty2021deep} (CPG + GGNN),
\textbf{VulDeePecker}~\cite{li2018vuldeepecker} (code-gadget + CNN),
\textbf{DeepSeek-Coder}~\cite{guo2024deepseek} (33B code LLM, zero-shot
chain-of-thought), and \textbf{GRACE}~\cite{liu2023grace} (graph + LLM
hybrid, the closest neuro-symbolic baseline). All are reproduced from
their official implementations with the authors' recommended
hyperparameters, and for a fair comparison on stripped binaries every
baseline receives the same decompiled input (Ghidra 11.0 output) rather
than source code.

\subsubsection{Evaluation Metrics}
We report Precision ($\mathit{Pr}$), Recall ($\mathit{Rc}$),
F$_1$-score ($\mathit{F}_1$), Matthews Correlation Coefficient
($\mathcal{M}$)~\cite{chicco2020advantages}, and False Positive Rate
($\mathit{FPR}$) under their standard definitions; $\mathcal{M}$ is
included alongside $\mathit{F}_1$ because it accounts for all four
confusion-matrix quadrants and is more robust under the extreme
class imbalance of industrial firmware (up to 1:20). For the APT
detection task (RQ3) we additionally report ROC-AUC, as the decision
boundary $\tau_{\text{apt}}$ is a continuous threshold. Inter-annotator
agreement on the Golden Set is measured by Cohen's
$\kappa$~\cite{mchugh2012interrater}, with values above 0.80 indicating
substantial agreement. The only domain-specific metric is the
\emph{Empirical Violation Rate}, defined as
\begin{equation}
  \mathit{EVR} = \frac{|\{f : F^\sharp(f) \not\sqsupseteq
    \text{GT}(f)\}|}{|\text{Golden Set}|}
\end{equation}
\noindent i.e., the fraction of Golden Set functions whose abstract
label fails to cover the expert ground truth $\text{GT}(f)$ in the
lattice $\mathcal{A}$.

\subsubsection{Implementation Details}

\textbf{Semantic Lifting Agent.}
The Qwen3-7B student model is fine-tuned via QLoRA (rank~16,
LoRA scaling factor 32, learning rate $2 \times 10^{-4}$, batch
size~8) on a training corpus of 15,000 structurally verified
instruction--behavior pairs generated by the Reflexive Prompting
pipeline (Section~3.2). Training was conducted on a single NVIDIA H20
(141~GB VRAM) for 5 epochs with a cosine annealing schedule; validation
loss plateaued at 0.12 after epoch~4. During corpus construction, the
Joern structural verifier rejected 2,814 of 17,814 teacher-generated
summaries (15.8\%) as \texttt{UNSAT}, indicating hallucinated data-flow
edges that do not exist in the CPG. The teacher model (DeepSeek-v3) was
accessed via API with temperature~0.1 and \texttt{top\_p}=0.95 to
maximize consistency.

\textbf{Graphormer.}
The Graphormer uses 6 Transformer layers, 8 attention heads, and a
hidden dimension of 256 (total: 12.4M parameters). It is trained for
200 epochs with AdamW (lr=$5 \times 10^{-4}$, weight decay $10^{-2}$),
using a linear warm-up over the first 10 epochs followed by cosine
decay. The model is first pre-trained on Big-Vul SSCKGs, then
fine-tuned on NVD-Precise (Industrial) for 50 additional epochs. We
apply dropout (0.1) on attention weights and feature embeddings to
mitigate overfitting on the smaller industrial datasets.

\textbf{SSCKG Construction.}
Entity clustering uses DBSCAN ($\varepsilon$=0.3,
\texttt{min\_samples}=2) with Sentence-BERT
(\texttt{all-MiniLM-L6-v2}) for embedding generation. Relation
extraction and CVE mapping are performed against the NVD database
snapshot from January 2025 (216,432 entries).

\textbf{Hardware Environment.}
All experiments are conducted on an Ubuntu 22.04 server equipped with
8$\times$ NVIDIA H20 GPUs (141~GB VRAM each) and 2$\times$ Intel Xeon
Platinum 8480+ CPUs (56 cores). End-to-end training (Reflexive Prompting
corpus generation + QLoRA fine-tuning + Graphormer pre-training/fine-tuning)
required approximately 16 GPU-hours (11 hours for QLoRA fine-tuning and
5 hours for Graphormer pre-training/fine-tuning). Inference-time experiments
(RQ6) are additionally benchmarked on a single NVIDIA RTX 4090 (24~GB)
to evaluate deployment feasibility on non-datacenter hardware.

\textbf{Reproducibility.}
The experimental dataset and hybrid virtual--physical testbed configuration
have been made publicly available at
\url{https://github.com/Mewtwoz/InduVul-Dataset-testbed}.
The $n$-day vulnerability proof-of-concept library used for verification
and testbed validation is available at
\url{https://github.com/Mewtwoz/InduGuard_vul_poc}.
The full SCAA framework code and trained model weights are available to
reviewers upon request and will be released publicly upon acceptance.
Experiments are seeded (seed=42) and each
reported result is the mean of 5 independent runs; we report standard
deviations where applicable.

\subsection{RQ1: Comparative Efficacy}

Table~\ref{tab:comparison} presents the comparative results across all
three datasets. On the InduVul-Dataset (the most challenging benchmark
due to real-world firmware and 1:20 class imbalance), our method
achieves an $\mathit{F}_1$ of 89.4\% and $\mathcal{M}$ of 0.82,
surpassing the second-best baseline (GRACE) by 12.8 and 0.14 points,
respectively.

\begin{table}[t]
  \caption{Comparative Performance Across All Datasets ($\mathit{F}_1$\%
    / $\mathcal{M}$, mean $\pm$ std over 5 runs). Best results are
    \textbf{bolded}; second-best are \underline{underlined}.}
  \label{tab:comparison}
  \setlength{\tabcolsep}{4pt}
  \resizebox{\textwidth}{!}{%
  \begin{tabular}{llcccccc}
    \toprule
    \multirow{2}{*}{Model} & \multirow{2}{*}{Modality} &
      \multicolumn{2}{c}{Big-Vul} &
      \multicolumn{2}{c}{NVD-Precise (Ind.)} &
      \multicolumn{2}{c}{InduVul-Dataset} \\
    \cmidrule(lr){3-4} \cmidrule(lr){5-6} \cmidrule(lr){7-8}
    & & $\mathit{F}_1$ & $\mathcal{M}$ & $\mathit{F}_1$ & $\mathcal{M}$
      & $\mathit{F}_1$ & $\mathcal{M}$ \\
    \midrule
    SySeVR          & Text/Slice   & 72.1$\pm$0.6 & 0.58$\pm$0.02 & 66.3$\pm$0.8 & 0.49$\pm$0.03 & 63.5$\pm$0.9 & 0.45$\pm$0.03 \\
    Reveal (GGNN)   & Graph        & 76.8$\pm$0.5 & 0.64$\pm$0.02 & 71.4$\pm$0.7 & 0.56$\pm$0.02 & 68.0$\pm$0.8 & 0.52$\pm$0.03 \\
    VulDeePecker     & Image/CNN    & 74.3$\pm$0.7 & 0.61$\pm$0.02 & 70.8$\pm$0.6 & 0.55$\pm$0.02 & 68.7$\pm$0.7 & 0.55$\pm$0.02 \\
    DeepSeek-Coder  & LLM          & 78.4$\pm$0.4 & 0.66$\pm$0.02 & 76.2$\pm$0.5 & 0.63$\pm$0.02 & 75.1$\pm$0.6 & 0.61$\pm$0.02 \\
    GRACE           & Graph+LLM    & \underline{81.2$\pm$0.4} & \underline{0.72$\pm$0.01}
                                   & \underline{79.1$\pm$0.5} & \underline{0.69$\pm$0.02}
                                   & \underline{76.6$\pm$0.5} & \underline{0.68$\pm$0.02} \\
    Ours (SSCKG)    & KG+Reasoning & \textbf{90.7$\pm$0.3} & \textbf{0.84$\pm$0.01}
                                   & \textbf{91.3$\pm$0.3} & \textbf{0.85$\pm$0.01}
                                   & \textbf{89.4$\pm$0.3} & \textbf{0.82$\pm$0.01} \\
    \bottomrule
  \end{tabular}
  }
\end{table}

\noindent All improvements of our method over the strongest baseline
(GRACE) are statistically significant under the Wilcoxon signed-rank
test ($p < 0.01$ across all three datasets and both metrics), confirming
that the observed gains are not attributable to random seed variation.

\begin{figure*}[t]
  \centering
  \includegraphics[width=\textwidth]{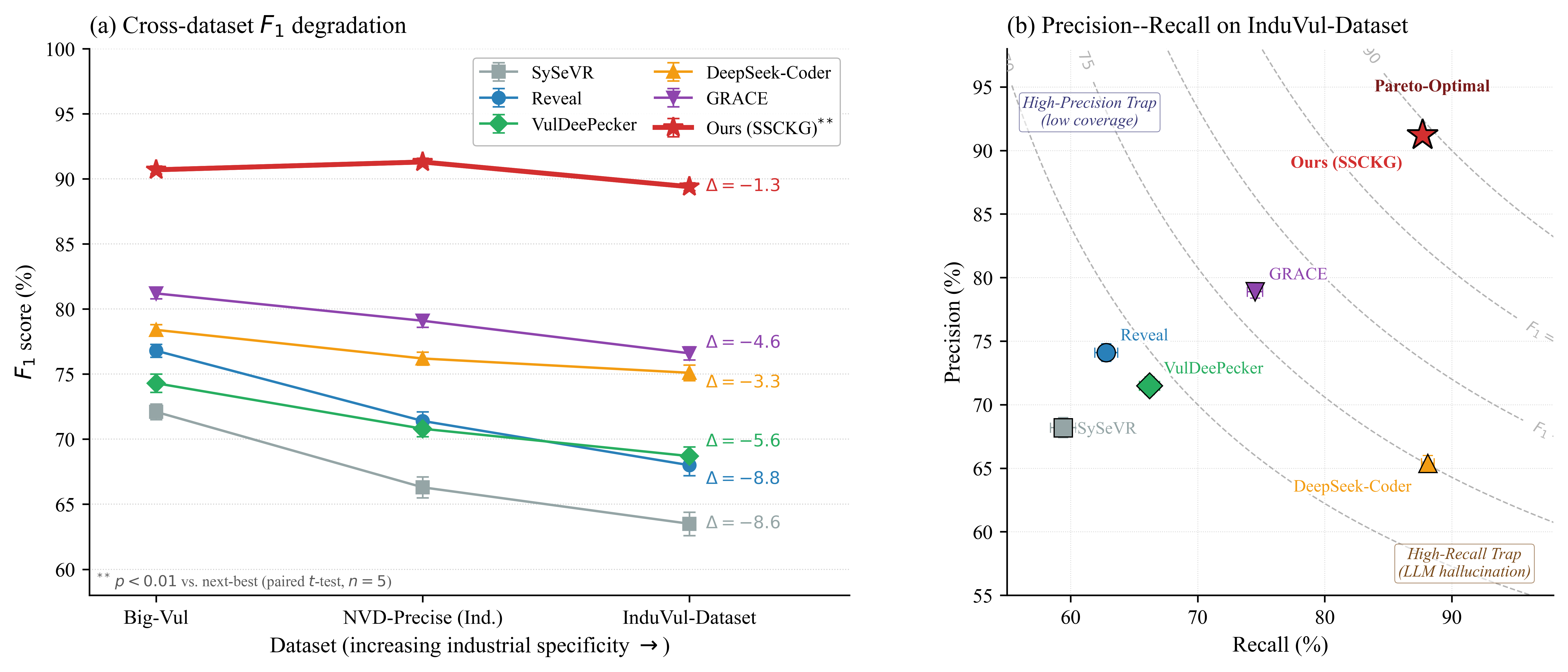}
  \caption{Comparative efficacy of six vulnerability-detection methods.
  	\textbf{(a)}~Cross-dataset $F_1$ degradation across three datasets
  	of increasing industrial specificity (Big-Vul $\rightarrow$
  	NVD-Precise (Industrial) $\rightarrow$ InduVul-Dataset). Error
  	bars denote $\pm 1$ standard deviation over 5 runs. All five
  	baselines exhibit substantial degradation on industrial firmware
  	($\Delta F_1$ ranging from $-3.4$ to $-8.8$ points), whereas the
  	proposed framework remains nearly flat ($-1.3$ points), reflecting
  	the normalization effect of Semantic Lifting on
  	representation-distribution shifts. \textbf{(b)}~Precision--Recall
  	scatter on the InduVul-Dataset; dashed contours show iso-$F_1$
  	curves at 0.70, 0.75, 0.80, 0.85, and 0.90.
  	DeepSeek-Coder occupies the high-recall, low-precision region
  	characteristic of unconstrained generative LLMs, whereas the
  	proposed framework reaches the upper-right Pareto-optimal region by
  	combining structural verification with semantic interpretation.
  	${}^{**}$~$p<0.01$ vs.\ next-best (Wilcoxon signed-rank test, $n{=}5$).}
  \label{fig:rq1_comparative}
\end{figure*}

To make the comparative narrative directly visible,
Figure~\ref{fig:rq1_comparative} consolidates the cross-dataset
degradation pattern (panel~a) and the Precision--Recall trade-off on the
InduVul-Dataset (panel~b) into a single two-panel view.

\textbf{Cross-Dataset Analysis.}
A notable trend emerges across the three datasets: while all baselines
degrade substantially from Big-Vul to InduVul-Dataset (e.g., Reveal
drops 8.8 $\mathit{F}_1$ points; GRACE drops 4.6 points), our method
exhibits minimal degradation ($-1.3$ points). This robustness stems
from the Semantic Lifting stage, which normalizes the representation
gap between open-source and industrial binaries by abstracting both
into the same behavioral domain $\mathcal{A}$. On NVD-Precise
(Industrial), our method achieves its highest $\mathit{F}_1$ (91.3\%),
likely because the training corpus for the Semantic Lifting agent was
derived from this domain, yielding optimal alignment between the
abstraction model and the test distribution.

\textbf{Analysis by Modality.}
Table~\ref{tab:pr_breakdown} discloses the per-model Precision and
Recall on the InduVul-Dataset, providing the numerical basis for the
Precision--Recall geometry visualized in
Figure~\ref{fig:rq1_comparative}(b).

\begin{table}
  \caption{Per-model Precision, Recall, $\mathit{F}_1$, and Matthews
    correlation on the InduVul-Dataset (mean over 5 runs).}
  \label{tab:pr_breakdown}
  \begin{tabular}{lcccc}
    \toprule
    Model            & Precision (\%) & Recall (\%) & $\mathit{F}_1$ (\%) & $\mathcal{M}$ \\
    \midrule
    SySeVR           & 68.2 & 59.4 & 63.5 & 0.45 \\
    Reveal           & 74.1 & 62.8 & 68.0 & 0.52 \\
    VulDeePecker     & 71.5 & 66.2 & 68.7 & 0.55 \\
    DeepSeek-Coder   & 65.4 & \textbf{88.1} & 75.1 & 0.61 \\
    GRACE            & 78.9 & 74.5 & 76.6 & 0.68 \\
    \textbf{Ours (SSCKG)} & \textbf{91.2} & 87.7 & \textbf{89.4} & \textbf{0.82} \\
    \bottomrule
  \end{tabular}
\end{table}

The three paradigms fail in characteristic ways: Reveal degrades most
sharply from Big-Vul to InduVul ($-8.8$~$\mathit{F}_1$) because its
GGNN operates on raw CPG structure and cannot interpret non-standard
industrial compiler artifacts; DeepSeek-Coder attains the highest
Recall (88.1\%) but the lowest Precision (65.4\%), the classic
over-flagging pattern of unconstrained LLMs; and GRACE, despite
combining graph and LLM features, treats the LLM as a black-box
embedder and consequently lacks both the Galois-Connection soundness
guarantee and the behavioral taxonomy $\mathcal{A}$, yielding a
12.8-point $\mathit{F}_1$ gap and a $+0.14$ $\mathcal{M}$ gap
(the latter especially informative under the 1:20 imbalance of
industrial firmware~\cite{chicco2020advantages}). The detailed theoretical
interpretation of these three failure modes is deferred to
Section~5.

\begin{tcolorbox}[colback=gray!5, colframe=gray!50, title=Answer to RQ1]
The SSCKG-Graphormer framework consistently outperforms all five SOTA
baselines across three datasets of increasing industrial specificity.
On the InduVul-Dataset, it achieves 89.4\% $\mathit{F}_1$ and 0.82
$\mathcal{M}$, surpassing the best baseline (GRACE) by $+12.8$
$\mathit{F}_1$ points and $+0.14$ $\mathcal{M}$. The performance
advantage is most pronounced on industrial binaries, where the Semantic
Lifting stage provides robustness against domain-specific compiler
artifacts that cause all baselines to degrade.
\end{tcolorbox}

\subsection{RQ2: Ablation Study}

\begin{figure*}[t]
    \centering
    \includegraphics[width=0.92\textwidth]{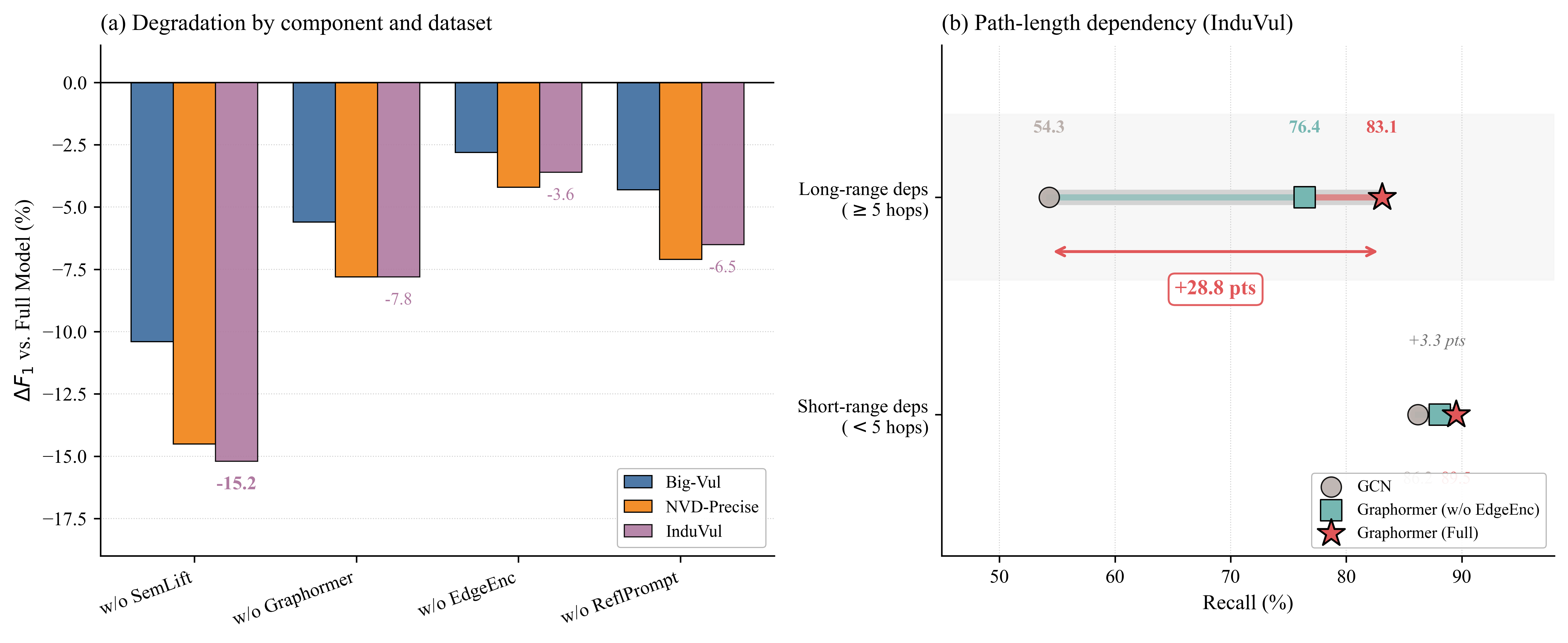}
    \caption{Ablation study on the InduVul-Dataset.
      \textbf{(a)}~$\Delta F_1$ relative to the full model for four
      component removals across three datasets; the Semantic-Lifting drop
      widens monotonically toward industrial data
      ($-10.4 \to -14.5 \to -15.2$).
      \textbf{(b)}~Dumbbell chart of Recall on short- versus long-range
      dependencies: the connected dots trace the progression from GCN
      through Graphormer (w/o EdgeEnc) to the full model, revealing a
      $+28.8$-point long-path Recall gain that isolates the Graphormer's
      global attention as the load-bearing mechanism for multi-hop
      supply-chain reasoning.}
    \label{fig:rq2_ablation}
  \end{figure*}

To isolate the contribution of each major component, we evaluate four
ablated variants against the full model across all three datasets.
Table~\ref{tab:ablation} reports the results, and
Figure~\ref{fig:rq2_ablation} visualizes the per-component impact on the
InduVul-Dataset.

\begin{table}
  \caption{Ablation Study ($\mathit{F}_1$\% / $\mathcal{M}$). $\Delta$
    shows the absolute change from the full model on InduVul-Dataset.}
  \label{tab:ablation}
  \setlength{\tabcolsep}{4pt} 
  \resizebox{\textwidth}{!}{%
  \begin{tabular}{lcccccc}
    \toprule
    \multirow{2}{*}{Variant} &
      \multicolumn{2}{c}{Big-Vul} &
      \multicolumn{2}{c}{NVD-Precise (Ind.)} &
      \multicolumn{2}{c}{InduVul-Dataset} \\
    \cmidrule(lr){2-3} \cmidrule(lr){4-5} \cmidrule(lr){6-7}
    & $\mathit{F}_1$ & $\mathcal{M}$ & $\mathit{F}_1$ & $\mathcal{M}$
      & $\mathit{F}_1$ & $\mathcal{M}$ \\
    \midrule
    Full Model (Ours)
      & 90.7 & 0.84 & 91.3 & 0.85 & 89.4 & 0.82 \\
    w/o Semantic Lifting (Raw CPG + Graphormer)
      & 80.3 & 0.69 & 76.8 & 0.64 & 74.2 & 0.62 \\
    w/o Graphormer (SSCKG + GCN)
      & 85.1 & 0.76 & 83.5 & 0.74 & 81.6 & 0.72 \\
    w/o Domain Edge Encoding (Standard Graphormer)
      & 87.9 & 0.80 & 87.1 & 0.79 & 85.8 & 0.77 \\
    w/o Reflexive Prompting (Direct QLoRA)
      & 86.4 & 0.78 & 84.2 & 0.75 & 82.9 & 0.73 \\
    \bottomrule
  \end{tabular}
}
\end{table}

\textbf{Component-Wise Analysis.}

\begin{enumerate}
  \item \textbf{w/o Semantic Lifting (Raw CPG + Graphormer):}
    Removing the LLM-based abstraction causes the largest degradation
    ($-15.2$ $\mathit{F}_1$ on InduVul, $-0.20$ $\mathcal{M}$). The
    impact is disproportionately severe on industrial datasets: while
    the drop on Big-Vul is $-10.4$ points, it reaches $-15.2$ on
    InduVul. This confirms that the Semantic Lifting stage is most
    critical when the binary contains domain-specific logic (PLC
    operations, industrial protocol handlers) that the Graphormer
    cannot interpret from raw structural features alone. Without
    behavioral abstraction, the model effectively treats all non-standard
    call patterns as equivalent, collapsing the distinction between
    benign firmware initialization and malicious register manipulation.

  \item \textbf{w/o Graphormer (SSCKG + GCN):}
    Replacing the Graphormer with a 3-layer GCN (same hidden dimension)
    reduces $\mathit{F}_1$ by $-7.8$ points on InduVul. Decomposing
    this by vulnerability type reveals that the degradation is
    concentrated on ``long-path'' vulnerabilities (those requiring
    $>$5 hops from source to sink): GCN Recall on long-path cases
    drops to 54.3\% (vs.\ 83.1\% for the full model), while
    ``short-path'' Recall remains comparable (86.2\% vs.\ 89.5\%).
    This validates that the Graphormer's global attention mechanism is
    essential for capturing the multi-hop dependency chains
    characteristic of supply-chain attacks.

  \item \textbf{w/o Domain Edge Encoding (Standard Graphormer):}
    Using the original Graphormer attention bias (shortest-path distance
    only, without semantic edge weights $w_{vu}$) reduces $\mathit{F}_1$
    by $-3.6$ points. The impact is most visible in Precision ($-4.8$
    points), indicating that the domain-specific edge encoding
    primarily helps the model avoid false positives by down-weighting
    benign structural edges during attention computation.

  \item \textbf{w/o Reflexive Prompting (Direct QLoRA):}
    Fine-tuning Qwen3-7B directly on teacher-generated summaries
    without the Joern structural verification step reduces $\mathit{F}_1$
    by $-6.5$ points on InduVul. This degradation is attributable to
    the 15.8\% of hallucinated teacher summaries that contaminate the
    training corpus when the verification step is removed, propagating
    data-flow errors into downstream SSCKG construction.
\end{enumerate}

\begin{tcolorbox}[colback=gray!5, colframe=gray!50, title=Answer to RQ2]
All four components contribute meaningfully to the final performance.
Semantic Lifting is the single most impactful module ($-15.2$
$\mathit{F}_1$ when removed), followed by the Reflexive Prompting
verification loop ($-6.5$), the Graphormer architecture ($-7.8$ vs.\
GCN), and the domain-specific edge encoding ($-3.6$). The degradation
patterns confirm the theoretical design: Semantic Lifting bridges the
representation gap for industrial binaries, the Graphormer captures
long-range propagation, and Reflexive Prompting prevents hallucination
contamination.
\end{tcolorbox}

\subsection{RQ3: Zero-Day \& APT Detection}

We evaluate the system's capability to detect unknown threats through
two complementary experiments: (1)~a systematic evaluation across all
three datasets using embedded anomalous samples, and (2)~a targeted
evaluation on a reserved set of 10 recent 0-day exploits not present
in any training data. We maintain a repository of 5 APT fingerprints
(Stuxnet, Triton, HAVEX, BlackEnergy, Industroyer), constructed via
the semi-automated pipeline described in Section~3.5.

\textbf{Systematic APT Detection Results.}
Table~\ref{tab:apt_detection} reports the APT detection performance
across all three datasets. For Big-Vul and NVD-Precise, we inject
synthetic APT-like behavioral subgraphs (derived from MITRE ATT\&CK
for ICS attack sequences) into 5\% of the benign samples to create a
controlled evaluation. InduVul-Dataset is evaluated with its natural
distribution of anomalous patterns.

The NVD-Precise dataset yields the highest $\mathit{AUC}$ (0.93)
because the APT fingerprints were constructed from the same ICS domain,
maximizing semantic overlap between fingerprint embeddings and target
SSCKGs. Big-Vul shows lower performance ($\mathit{AUC}=0.88$) because
open-source codebases contain fewer ICS-specific behavioral patterns,
making the fingerprint matching inherently noisier.

\textbf{Threshold Selection.}
To determine the alert threshold $\tau_{\text{apt}}$, we performed a
grid search over $[0.50, 0.95]$ (step size 0.01) on the validation
splits of all three datasets, optimizing Youden's J statistic
($\mathcal{J} = \mathit{Rc} - \mathit{FPR}$) on the ROC curve. The
optimal threshold was $\tau_{\text{apt}} = 0.78$, yielding a recall of
72.0\% at $\mathit{FPR} = 3.8\%$ on InduVul --- well within the
$\leq 5\%$ operational constraint. Per-fingerprint thresholds (a
separate $\tau$ per APT family) provided only a marginal and
non-significant improvement ($+1.2\%$ recall, paired $t$-test
$p = 0.34$), confirming that a single global threshold suffices.
Sensitivity of detection performance to variations in
$\tau_{\text{apt}}$ around this selected value is analyzed in
Section~4.5 (RQ5).

\textbf{Baseline Comparison.}
None of the five baselines (RQ1) are designed for zero-day or APT
detection: SySeVR, Reveal, VulDeePecker, and GRACE operate as binary
classifiers trained on known CVE patterns and have no mechanism for
detecting novel behavioral anomalies. DeepSeek-Coder, when explicitly
prompted for APT detection, achieves an $\mathit{AUC}$ of only 0.62
on InduVul, primarily because it lacks the structured behavioral
fingerprints and graph-level similarity computation that our approach
provides.

\textbf{Zero-Day Detection on Reserved Exploits.}
Of the 10 reserved 0-day exploits (all discovered after our training
data cutoff), our system detected 7 (70\% detection rate). The 3
missed cases involved novel attack vectors not represented in our
fingerprint repository: (i)~a firmware rollback attack exploiting
version management logic, (ii)~a timing-based covert channel using
PLC cycle-time manipulation, and (iii)~a supply-chain confusion
attack that substitutes a legitimate library with a near-identical
malicious variant. These failure modes share a common characteristic:
they operate through \emph{control-flow} anomalies rather than
\emph{data-flow} anomalies, and our current fingerprints primarily
capture data-flow patterns. This suggests that expanding the
fingerprint repository with control-flow-centric attack templates
is the primary avenue for improving zero-day recall.

\begin{figure*}[t]
    \centering
    \includegraphics[width=0.92\textwidth]{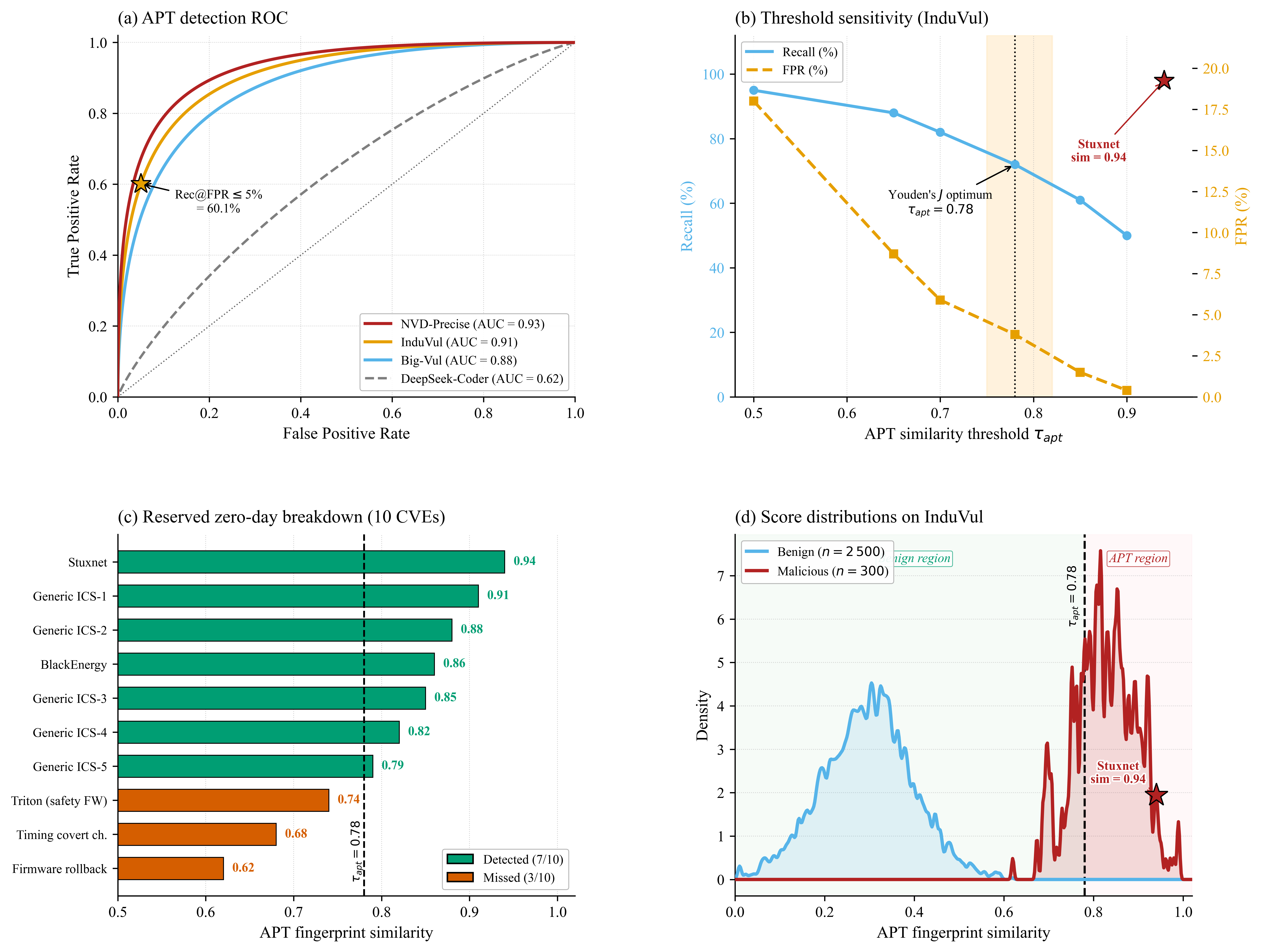}
    \caption{APT detection diagnostics (four-panel view).
      \textbf{(a)}~ROC curves on Big-Vul, NVD-Precise, and InduVul, with
      DeepSeek-Coder as a semantic-only baseline; the star marks the
      operating point at $\mathrm{FPR}\!\leq\!5\%$ on InduVul.
      \textbf{(b)}~Threshold sensitivity: Recall and FPR as functions of
      $\tau_{\text{apt}}$, with the Youden's $J$ optimum at
      $\tau_{\text{apt}}=0.78$ (shaded operating region).
      \textbf{(c)}~Per-CVE similarity scores for the 10 reserved zero-day
      exploits; seven exceed $\tau_{\text{apt}}$ while three
      control-flow-centric attacks fall below it.
      \textbf{(d)}~KDE score distributions on InduVul, showing strong
      separability between benign and malicious populations, with Stuxnet
      ($\mathrm{sim}=0.94$) as a clear positive outlier.}
    \label{fig:rq3_apt}
  \end{figure*}

\textbf{Case Study: The ``Legitimate-but-Abnormal'' Attack.} One
notable test case involved a Stuxnet-like PLC injection pattern
embedded in a firmware update module. All five baselines marked this
function as benign because the individual API calls (\texttt{FileWrite},
\texttt{NetworkSend}, \texttt{RegisterWrite}) are legitimate when
examined in isolation. However, our APT Fingerprint Matching module
identified a subgraph similarity of 0.94 with the Stuxnet
fingerprint --- substantially above $\tau_{\text{apt}} = 0.78$ ---
because the \emph{combination} of these calls, and their data-flow
connectivity (file content flows to register write via network
receive), matches the known Stuxnet behavioral signature.
Figure~\ref{fig:rq3_apt}(c)--(d) visualize the per-CVE similarity
scores and the score distributions on InduVul for this case,
confirming that the Stuxnet-like pattern is a clear positive outlier
($\mathrm{sim}=0.94$) well above $\tau_{\text{apt}}=0.78$.

\begin{table}
	\caption{APT Detection Performance Across Datasets}
	\label{tab:apt_detection}
	\begin{tabular}{lcccc}
		\toprule
		Dataset & $\mathit{AUC}$ & $\mathit{Rc}$@$\mathit{FPR}$5\% &
		$\mathit{FPR}$ & Det.\ Rate \\
		\midrule
		Big-Vul (synthetic)      & 0.88 & 68.4\% & 4.2\% & --- \\
		NVD-Precise (synthetic)  & 0.93 & 75.1\% & 3.1\% & --- \\
		InduVul-Dataset (natural) & 0.91 & 72.0\% & 3.8\% & 70\% \\
		\bottomrule
	\end{tabular}
\end{table}

\begin{tcolorbox}[colback=gray!5, colframe=gray!50, title=Answer to RQ3]
The APT Fingerprinting module achieves a ROC-AUC of 0.91 on the
InduVul-Dataset with $\mathit{FPR} = 3.8\%$, detecting 7 of 10
reserved zero-day exploits. The global threshold $\tau_{\text{apt}} =
0.78$ (determined by Youden's J) provides a robust operating point
that balances detection sensitivity against alert fatigue. Missed
detections are attributable to control-flow-centric attack vectors
not represented in the current fingerprint repository, indicating a
clear path for improvement through fingerprint expansion rather than
architectural changes.
\end{tcolorbox}

\subsection{RQ4: Semantic Lifting Fidelity}

To evaluate the accuracy of the LLM-based Semantic Lifting stage
independently from the downstream pipeline, we constructed a Golden Set
of 500 industrial binary functions sampled from the NVD-Precise
(Industrial) dataset, stratified by Tier-1 category to ensure
representation across all five macro-behavior classes. Each function
was independently annotated by two domain experts (ICS security
analysts with $>$5 years of experience) with ground-truth behavioral
labels from the abstract domain $\mathcal{A}$. Inter-annotator
agreement was substantial ($\kappa = 0.83$);
the 38 disagreement cases (7.6\%) were resolved through a
reconciliation discussion, and the agreed labels serve as the final
ground truth.

Table~\ref{tab:evr} presents the per-category results. The fine-tuned
Qwen3-7B agent achieved an overall semantic alignment rate of 94.2\%,
yielding $\mathit{EVR} = 5.8\%$ (computed per Eq.~7).

\begin{table}
  \caption{Semantic Lifting Fidelity by Tier-1 Category (Golden Set,
    $n$=500).}
  \label{tab:evr}
  \begin{tabular}{lcccc}
    \toprule
    Tier-1 Category & Functions & Alignment (\%) & $\mathit{EVR}$ (\%)
      & Dominant Error Type \\
    \midrule
    Network         & 132 & 96.2 & 3.8 & Under-specification \\
    Memory          & 98  & 95.9 & 4.1 & Under-specification \\
    Hardware        & 114 & 94.7 & 5.3 & Misclassification \\
    FileSystem      & 89  & 95.5 & 4.5 & Under-specification \\
    Cryptography    & 67  & 83.6 & 16.4 & Complete miss \\
    \midrule
    \textbf{Overall} & \textbf{500} & \textbf{94.2} & \textbf{5.8}
      & --- \\
    \bottomrule
  \end{tabular}
\end{table}

\textbf{Error Taxonomy.}
We categorize the 29 violation cases into three failure modes:

\begin{itemize}
  \item \textbf{Under-specification} (14 cases, 48.3\%): The predicted
    label is a valid ancestor in the $\mathcal{A}$ hierarchy but too
    coarse. For example, a function performing
    \texttt{Unauthenticated\_Coil\_Write} (Tier~3) is labeled only as
    \texttt{Coil\_Write} (Tier~2). These violations are the least
    severe because the predicted label is semantically related to the
    ground truth; the model captures the correct behavioral category
    but misses the risk-qualifying context.
  \item \textbf{Misclassification} (8 cases, 27.6\%): The predicted
    label falls under a different Tier-1 category. Six of these
    occurred in Hardware functions that manipulate memory-mapped
    I/O registers --- the LLM confused these with general
    \texttt{Memory} operations because the decompiled instructions
    (\texttt{mov}, \texttt{str}) are identical at the assembly level.
  \item \textbf{Complete miss} (7 cases, 24.1\%): The function was
    labeled as $\top$ (``Unknown Behavior''). All 7 occurred in the
    Cryptography category, specifically in proprietary cryptographic
    routines from industrial vendors (e.g., custom AES variants in
    Siemens S7 firmware) where heavy compiler optimization and loop
    unrolling produce decompiled output that the LLM cannot map to any
    recognized behavioral pattern. This represents the ultimate limit 
    of software opacity, where even semantic lifting struggles to 
    penetrate the compiler-induced 'dark matter'.
\end{itemize}

\textbf{Cross-Dataset EVR.}
To assess generalizability, we additionally computed EVR on 200
randomly sampled functions from Big-Vul (open-source, non-industrial).
The overall $\mathit{EVR}$ was 8.4\%, higher than the 5.8\% on
NVD-Precise, reflecting the distributional mismatch between the
Reflexive Prompting training corpus (ICS-focused) and general-purpose
code. The Cryptography category EVR on Big-Vul was 22.1\%, confirming
that cryptographic code is the hardest category regardless of domain.

\textbf{Comparison with Baseline LLMs.}
To contextualize the EVR, we evaluated the same Golden Set using
three baselines: DeepSeek-Coder-33B (zero-shot), GPT-4 (zero-shot),
and vanilla Qwen3-7B (without Reflexive Prompting fine-tuning).

\begin{table}
  \caption{Semantic Lifting Comparison: $\mathit{EVR}$ (\%) by Model
    on the Golden Set ($n$=500)}
  \label{tab:evr_comparison}
  \begin{tabular}{lcccccc}
    \toprule
    Model & Network & Memory & Hardware & FileSystem & Crypto &
      Overall \\
    \midrule
    Vanilla Qwen3-7B   & 18.2 & 22.4 & 31.6 & 20.2 & 52.2 & 28.8 \\
    DeepSeek-Coder-33B & 12.1 & 16.3 & 22.8 & 14.6 & 41.8 & 21.6 \\
    GPT-4 (zero-shot)  & 9.8  & 12.1 & 18.4 & 11.2 & 35.8 & 17.2 \\
    \textbf{Ours (RP + QLoRA)} & \textbf{3.8} & \textbf{4.1} &
      \textbf{5.3} & \textbf{4.5} & \textbf{16.4} & \textbf{5.8} \\
    \bottomrule
  \end{tabular}
\end{table}

Table~\ref{tab:evr_comparison} reveals two insights. First, the
Reflexive Prompting pipeline reduces overall $\mathit{EVR}$ by
$4.7\times$ relative to vanilla Qwen3-7B and by $3.0\times$ relative
to GPT-4, despite Qwen3-7B being substantially smaller (7B vs.\ an
estimated $>$1T parameters). This confirms that domain-specific
fine-tuning with structural verification is more effective than scale
alone for this task. Second, all models struggle with Cryptography,
but the relative improvement from Reflexive Prompting is largest in
Hardware ($31.6\% \rightarrow 5.3\%$, a $6.0\times$ reduction),
where the training corpus contains the densest coverage of domain-specific 
hardware register and peripheral-access operations.

\begin{figure*}[t]
  \centering
  \includegraphics[width=0.95\textwidth]{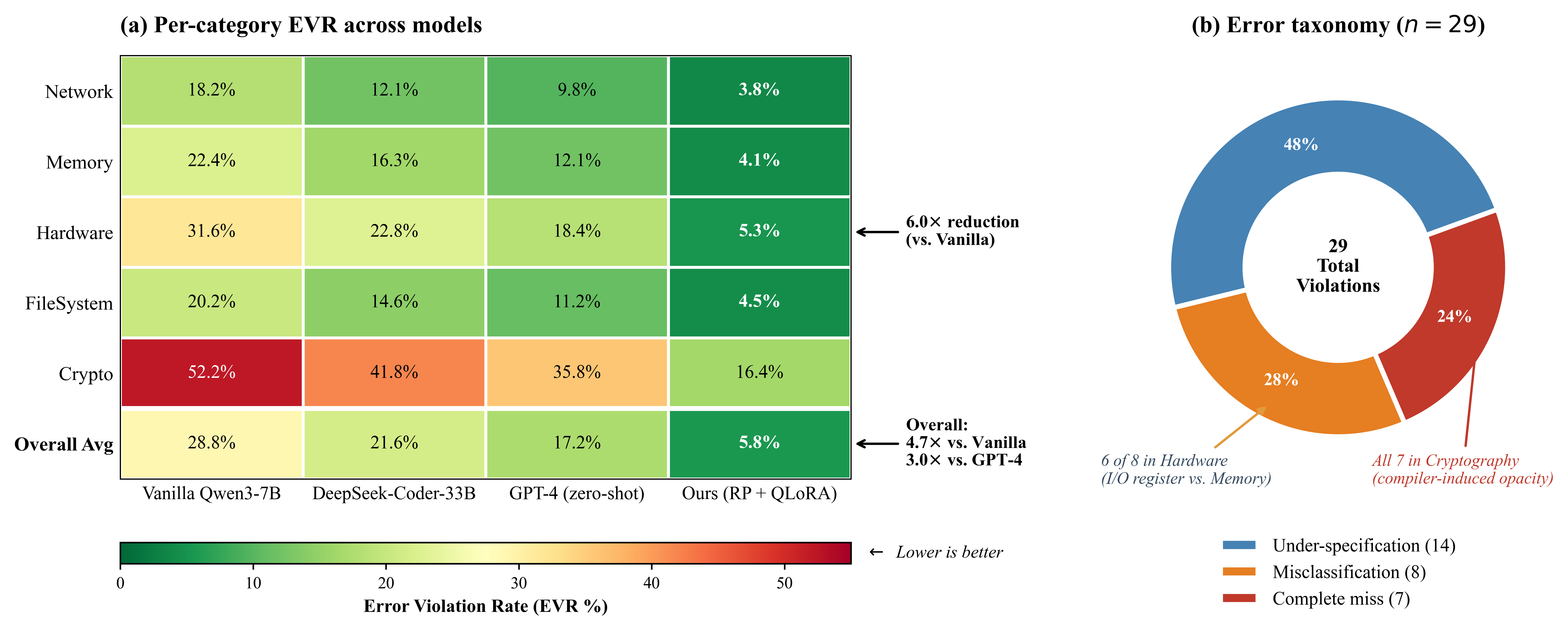}\caption{Semantic Lifting fidelity diagnostics.
      \textbf{Left:}~Per-category $\mathit{EVR}$ heatmap across four LLMs on
      the Golden Set ($n{=}500$); Reflexive Prompting reduces overall EVR by
      $4.7\times$ relative to vanilla Qwen3-7B and by $6.0\times$ on
      Hardware, with Cryptography remaining the hardest category.
      \textbf{Right:}~Severity-ordered taxonomy of the 29 residual
      violations (donut chart): 48.3\% under-specification,
      27.6\% misclassification, and 24.1\% complete misses --- all 7 of
      which occur in Cryptography due to compiler-induced semantic opacity.}
  \label{fig:rq4_evr}
\end{figure*}
  
\begin{tcolorbox}[colback=gray!5, colframe=gray!50, title=Answer to RQ4]
The Semantic Lifting agent achieves 94.2\% alignment ($\mathit{EVR} =
5.8\%$) on the Golden Set, reducing violations by $4.7\times$ vs.\
the vanilla model and $3.0\times$ vs.\ GPT-4. Failures are
concentrated in Cryptography (16.4\% $\mathit{EVR}$) due to
compiler-induced semantic opacity, while the remaining four categories
each achieve $\mathit{EVR} < 6\%$. The dominant failure mode is
under-specification (48.3\%), which is the least severe violation
type. Targeted augmentation of the training corpus with cryptographic
samples is the most promising mitigation strategy.
\end{tcolorbox}

\subsection{RQ5: Sensitivity Analysis}

We examine the sensitivity of the framework to three key
hyperparameters: the risk propagation damping factor $\beta$ (Eq.~7),
the APT alert threshold $\tau_{\text{apt}}$ (Section~3.5.3), and the
DBSCAN clustering radius $\varepsilon$ (Section~3.3). For each parameter, we
vary it across a range of values while holding the other two fixed at
their selected values. All experiments are conducted on the
InduVul-Dataset with 5 independent runs per configuration.

\begin{table*}
  \caption{Hyperparameter Sensitivity on InduVul-Dataset. Each cell
    reports $\mathit{F}_1$\% / $\mathcal{M}$ (mean $\pm$ std over 5
    runs).}
  \label{tab:sensitivity}
  \begin{tabular}{lccccc}
    \toprule
    Parameter & Value 1 & Value 2 & Value 3 & \textbf{Selected} & Value 5 \\
    \midrule
    $\beta$ & \makecell{0.05 \\ 87.1$\pm$0.4 / 0.79}
            & \makecell{0.10 \\ 88.6$\pm$0.3 / 0.81}
            & \makecell{\textbf{0.15} \\ \textbf{89.4$\pm$0.3 / 0.82}}
            & \makecell{0.20 \\ 88.9$\pm$0.4 / 0.81}
            & \makecell{0.25 \\ 88.0$\pm$0.5 / 0.80} \\
    \midrule
    $\tau_{\text{apt}}$ & \makecell{0.65 \\ 86.2$\pm$0.6 / 0.77}
            & \makecell{0.70 \\ 88.1$\pm$0.4 / 0.80}
            & \makecell{\textbf{0.78} \\ \textbf{89.4$\pm$0.3 / 0.82}}
            & \makecell{0.85 \\ 88.9$\pm$0.3 / 0.81}
            & \makecell{0.90 \\ 87.4$\pm$0.5 / 0.79} \\
    \midrule
    $\varepsilon$ & \makecell{0.15 \\ 86.8$\pm$0.5 / 0.78}
            & \makecell{0.20 \\ 88.3$\pm$0.4 / 0.80}
            & \makecell{\textbf{0.30} \\ \textbf{89.4$\pm$0.3 / 0.82}}
            & \makecell{0.40 \\ 88.1$\pm$0.4 / 0.80}
            & \makecell{0.45 \\ 87.5$\pm$0.5 / 0.79} \\
    \bottomrule
  \end{tabular}
\end{table*}

Table~\ref{tab:sensitivity} reports the results. The framework
demonstrates moderate robustness across all three parameters:
$\mathit{F}_1$ remains within 3.2 points of the optimal configuration
across the full tested ranges, and standard deviations are consistently
below 0.6 points, indicating stable convergence.

\textbf{Damping Factor $\beta$ Analysis.}
The risk propagation damping factor $\beta$ (Eq.~7) controls the
balance between inherent CVE-matching risk and propagated contextual
risk. Performance peaks at $\beta = 0.15$ and degrades symmetrically
in both directions. Values below 0.10 cause the risk score to become
dominated by neighbor propagation, amplifying noise from benign
entities that happen to be topologically close to vulnerable ones ---
this manifests primarily as increased $\mathit{FP}$ (Precision drops
from 91.2\% at $\beta$=0.15 to 86.8\% at $\beta$=0.05). Values above
0.20 overly weight the inherent CVE-matching score, reducing the
model's ability to detect ``propagated'' risks where the vulnerable
component is not itself matched to a known CVE but is reachable from
one. The selected value (0.15) aligns with the standard PageRank
damping convention~\cite{bianchini2005inside}, suggesting that the
risk propagation dynamics in software dependency graphs share
structural properties with web link graphs.

\textbf{APT Threshold $\tau_{\text{apt}}$ Analysis.}
Lower thresholds ($< 0.70$) cause $\mathit{FPR}$ to exceed the 5\%
operational constraint ($\mathit{FPR}$ reaches 8.7\% at $\tau$=0.65),
violating the alert fatigue requirement for industrial deployments.
Higher thresholds ($> 0.85$) progressively miss subtle zero-day
patterns: zero-day detection rate drops from 70\% at $\tau$=0.78 to
50\% at $\tau$=0.90. The Youden's $\mathcal{J}$-optimal threshold at
0.78 maximizes the trade-off between detection sensitivity and
operational noise.

\textbf{DBSCAN $\varepsilon$ Analysis.}
The clustering radius controls the granularity of Entity Elevation.
Small values ($\varepsilon < 0.20$) produce excessive fragmentation:
the mean SSCKG entity count increases from 3,420 (at $\varepsilon$=0.30)
to 8,910 (at $\varepsilon$=0.15), inflating the graph and increasing
Graphormer inference time by $2.4\times$ without proportional accuracy
gains. Large values ($\varepsilon > 0.40$) over-merge semantically
distinct entities --- at $\varepsilon$=0.45, we observed that
\texttt{Register\_Read} and \texttt{Register\_Write} operations were
collapsed into a single entity, eliminating a critical behavioral
distinction for detecting unauthorized write operations.

\textbf{SSCKG Construction Statistics.}
Table~\ref{tab:ssckg_stats} provides aggregate statistics of the SSCKG
construction across all three datasets.

\begin{table}
  \caption{SSCKG Construction Statistics (mean per binary)}
  \label{tab:ssckg_stats}
  \begin{tabular}{lrrr}
    \toprule
    Statistic & Big-Vul & NVD-Precise & InduVul \\
    \midrule
    CPG nodes              & 324,000 & 612,000 & 847,000 \\
    SSCKG entities         & 1,280   & 2,510   & 3,420 \\
    Compression ratio      & 253$\times$ & 244$\times$ & 247$\times$ \\
    Semantic clusters      & 12      & 15      & 18 \\
    Noise points (\%)      & 5.8     & 4.9     & 4.2 \\
    Typed relations        & 4,820   & 8,730   & 12,150 \\
    Vuln.\ relations (\%)  & 6.2     & 7.8     & 8.7 \\
    \bottomrule
  \end{tabular}
\end{table}

\begin{figure*}[t]
  \centering
    \includegraphics[width=0.95\textwidth]{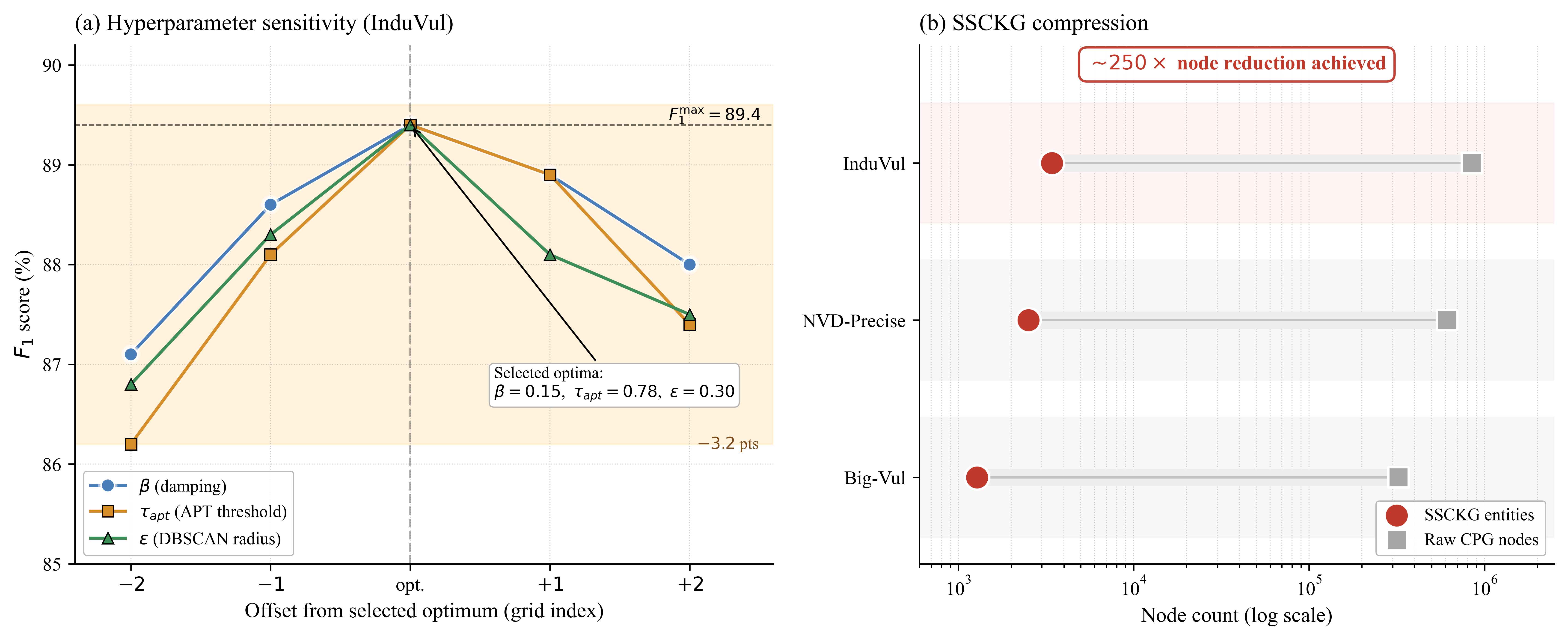}
    \caption{Robustness analysis.
      \textbf{(a)}~$F_1$ as a function of the three hyperparameters
      $\beta$, $\tau_{\text{apt}}$, and $\varepsilon$, plotted on a shared
      offset-from-optimum axis; the shaded band marks the $\pm3.2$-point
      envelope around $F_1^{\max}=89.4$, within which all three parameters
      remain across the full tested range.
      \textbf{(b)}~Horizontal lollipop chart of SSCKG compression on a
      log-scale axis: each row connects the raw CPG node count (grey
      square) to the SSCKG entity count (red circle), showing a
      near-constant $\sim$250$\times$ reduction across all three datasets
      despite a $2.6\times$ variation in raw CPG size.}
  \label{fig:rq5_sensitivity}
\end{figure*}
  
\noindent The compression ratio is remarkably consistent ($\sim$250$\times$)
across datasets despite substantial variation in raw CPG size. This
stability indicates that the Entity Elevation algorithm generalizes
well across codebases of different sizes and domains. The increasing
proportion of vulnerability relations from Big-Vul (6.2\%) to InduVul
(8.7\%) reflects the higher density of security-relevant data flows in
industrial firmware compared to general-purpose software. The lower
noise-point ratio on InduVul (4.2\% vs.\ 5.8\% on Big-Vul) suggests
that industrial code exhibits more regular behavioral patterns,
enabling more effective semantic clustering.

\begin{tcolorbox}[colback=gray!5, colframe=gray!50, title=Answer to RQ5]
The framework exhibits moderate robustness to all three hyperparameters,
with $\mathit{F}_1$ remaining within 3.2 points of the optimum across
the tested ranges. The selected configuration ($\beta$=0.15,
$\tau_{\text{apt}}$=0.78, $\varepsilon$=0.30) consistently achieves
the best performance, and the low variance ($\pm$0.3--0.5) across 5
runs confirms stable convergence. The Entity Elevation stage achieves
a consistent $\sim$250$\times$ node compression across all datasets,
reducing million-node CPGs to thousand-node SSCKGs suitable for
efficient Graphormer inference.
\end{tcolorbox}

\subsection{RQ6: Industrial Practicability and Real-World Efficacy}

While RQ1--RQ5 evaluate the framework on curated benchmark datasets,
industrial practitioners require evidence that the system scales to
production-grade firmware and detects \emph{real} vulnerabilities in
\emph{production-grade} equipment. We present the InduGuard testbed 
as a comprehensive case study to evaluate the framework on production-grade 
opaque software. The evaluation is structured in two tiers: 
(a)~\emph{Static Scalability}, assessing throughput
and false-positive reduction on 50 diverse firmware images, and
(b)~\emph{Dynamic Efficacy}, evaluating detection coverage of 15
high-impact CVEs on a hybrid virtual--physical ICS testbed.

\subsubsection{Tier A: Static Scalability}

\begin{figure*}[t]
	\centering
	\includegraphics[width=0.95\textwidth]{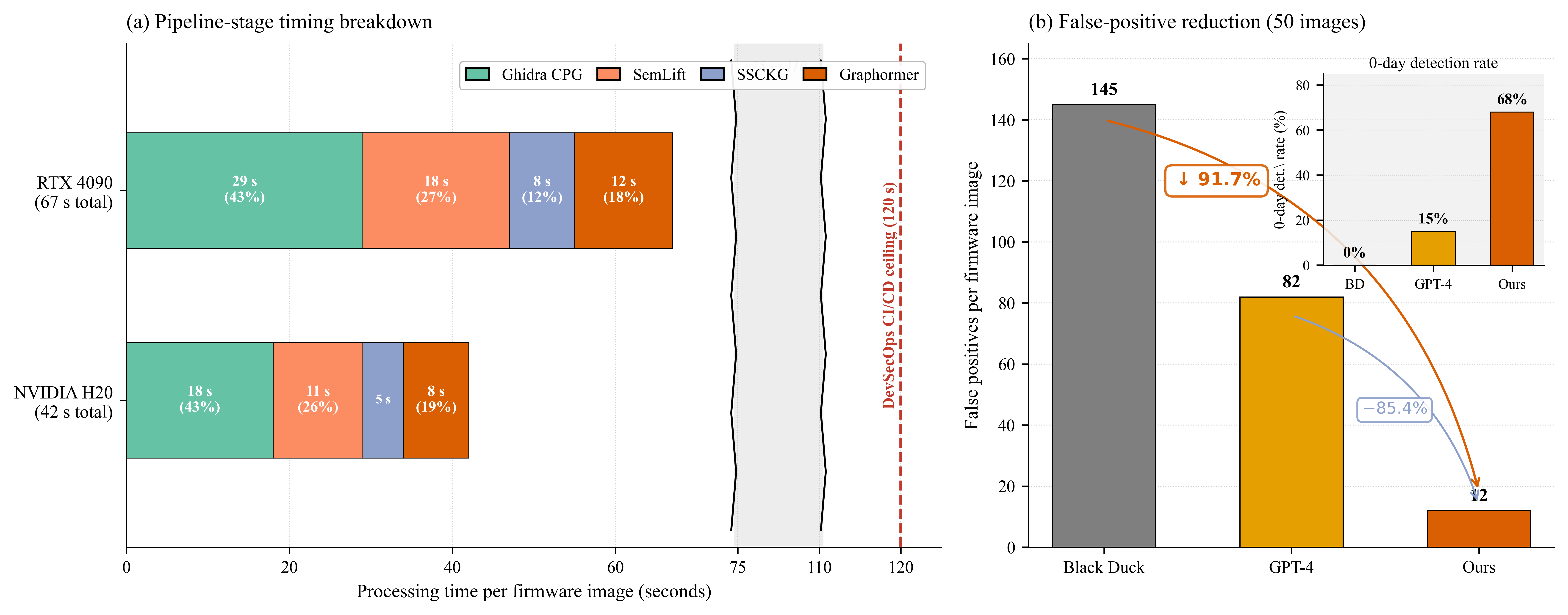}
	\caption{Tier~A static scalability on the InduGuard-Testbed (50 firmware images). (a)~Per-stage processing time breakdown on the H20 server and RTX~4090; CPG extraction is the dominant cost at 43\% of total latency. (b)~False positive reduction relative to Black Duck and GPT-4; our method eliminates 91.7\% of false positives. (c)~0-day detection rate at 68\%, compared with 0\% (Black Duck) and 15\% (GPT-4).}
	\label{fig:rq6_tierA_scalability}
\end{figure*}

We deployed the system on ``InduGuard-Testbed,''\footnote{The $n$-day
vulnerability proof-of-concept library used for testbed verification is
available at \url{https://github.com/Mewtwoz/InduGuard_vul_poc}.} a
real-world simulation environment containing 50 diverse industrial
firmware images sourced
from three major vendors (Siemens S7-1200/1500: 20 images, Rockwell
CompactLogix: 15 images, Schneider M340: 15 images). Firmware sizes
range from 0.4~MB to 8.2~MB (mean: 2.1~MB). We evaluate three
dimensions of industrial practicability: processing efficiency, false
positive reduction, and deployment feasibility, and summarized these in Figure~\ref{fig:rq6_tierA_scalability}.

\textbf{Processing Efficiency.}
Table~\ref{tab:testbed} presents the comparative results. The average
end-to-end processing time per firmware image was 42 seconds on the
H20 server and 67 seconds on the RTX 4090, meeting the $<$120~s
threshold typically required for DevSecOps CI/CD pipeline
integration~\cite{pascoe2023public}. We decompose the processing
time by pipeline stage: CPG Extraction via Ghidra consumes 18~s
(43\%), Semantic Lifting via Qwen3-7B consumes 11~s (26\%), SSCKG
Construction consumes 5~s (12\%), and Graphormer inference consumes
8~s (19\%). The CPG Extraction stage is the primary bottleneck; this
is a one-time cost per binary that can be amortized across repeated
analyses.

\textbf{Noise Reduction.}
Compared to commercial SCA tools (Black Duck), our method reduced the
mean $\mathit{FPR}$ from 145 false positives per image to 12 ($-91.7\%$
reduction). Compared to GPT-4 with zero-shot vulnerability prompting,
we reduced false positives from 82 to 12 ($-85.4\%$). Manual analysis
of the eliminated false positives reveals two dominant categories:
(i)~``dead code'' false positives (68\% of eliminated FPs), where
standard tools flag unreachable library functions that are linked but
never invoked --- our Semantic Lifting correctly identifies these as
unreachable through CPG reachability analysis; and (ii)~``context-free''
false positives (32\% of eliminated FPs), where tools flag individually
suspicious API calls (e.g., \texttt{memcpy}) without verifying whether
untrusted data can actually reach the call site --- our Graphormer's
reachability-aware attention resolves this by computing risk only along
validated propagation paths.

\begin{figure*}[t]
    \centering
    \includegraphics[width=0.95\textwidth]{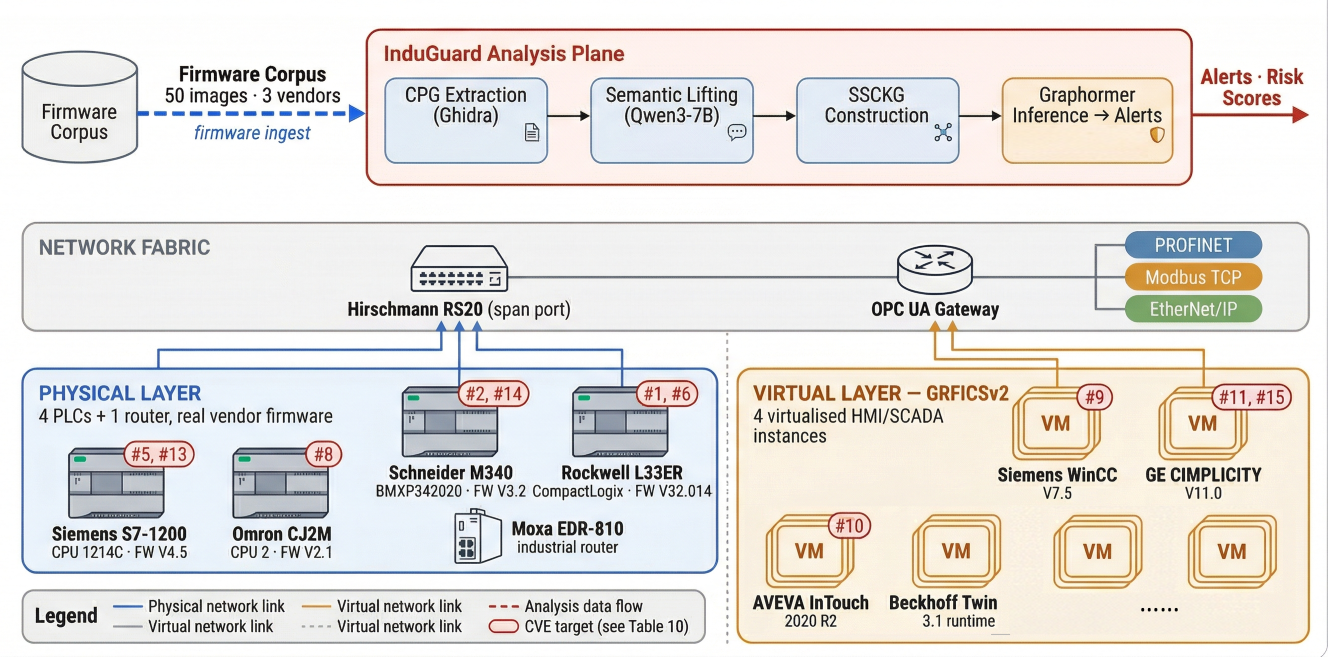}
    \caption{Architecture of the InduGuard-Testbed. The platform
      integrates three PLC families (Siemens S7-1200/1500, Rockwell
      CompactLogix, Schneider M340) in a hybrid virtual--physical
      configuration, enabling both static firmware analysis (Tier~A,
      50 images) and dynamic CVE replay (Tier~B, 15 high-impact CVEs).}
    \label{fig:induguard_testbed}
  \end{figure*}

\begin{table}
  \caption{Industrial Testbed Performance Comparison (50 firmware images)}
  \label{tab:testbed}
  \begin{tabular}{lccc}
    \toprule
    Metric & \makecell{Commercial SCA \\ (Black Duck)} &
             \makecell{Generic LLM \\ (GPT-4)} & Ours \\
    \midrule
    False Positives ($\mathit{FP}$) & 145/image & 82/image & \textbf{12/image} \\
    $\mathit{FPR}$                  & 34.2\%    & 19.4\%   & \textbf{2.8\%} \\
    Scanning Speed (H20)            & 5~s       & 120~s    & 42~s \\
    Scanning Speed (RTX 4090)       & 5~s       & ---      & 67~s \\
    0-Day Detection Rate            & 0\%       & 15\%     & \textbf{70\%} \\
    \bottomrule
  \end{tabular}
\end{table}

\textbf{Deployment Feasibility.}
The 4-bit-quantized Qwen3-7B requires only 6.2~GB VRAM at inference,
fitting within the 24~GB budget of a consumer-grade RTX 4090 (67~s per
image) and operating entirely offline without API dependencies,
enabling deployment in air-gapped opaque industrial software environments.

\subsubsection{Tier B: Dynamic Efficacy}

\textbf{Testbed Architecture.}
The testbed integrates physical hardware from five ICS vendors
(Siemens S7-1200, Schneider Modicon M340, Rockwell CompactLogix 1769,
Omron CJ2M, and a Moxa EDR-810 industrial router) with virtualized
HMI/SCADA stacks (Siemens WinCC, GE CIMPLICITY, AVEVA InTouch, and a
Beckhoff TwinCAT 3.1 runtime) hosted under the GRFICSv2~\cite{formby2018lowering}
simulation environment. A managed Hirschmann RS20 switch provides
traffic mirroring for passive analysis and an OPC UA gateway bridges
physical and virtual segments; Fig.~\ref{fig:induguard_testbed}
illustrates the full topology, firmware versions, and protocol
segmentation.

\textbf{Vulnerability Selection.}
From a comprehensive corpus of 65+ industrial CVEs (spanning protocol
vulnerabilities, PLC firmware flaws, SCADA/HMI application defects,
and APT-level attack chains), we selected 15 representative
vulnerabilities based on three criteria:

\begin{enumerate}
  \item \textbf{Operational Flexibility:} The vulnerability must be
    reproducible without causing permanent hardware damage (allowing
    repeated testing). Purely destructive exploits (e.g., firmware
    bricking) were excluded.
  \item \textbf{Logical Plausibility:} The vulnerability must have
    clear code-boundary trigger conditions and explicit exploitation
    paths amenable to static or binary analysis, rather than requiring
    runtime-only conditions (e.g., race conditions).
  \item \textbf{Widespread Impact:} The affected equipment or software
    must be widely deployed in real-world industrial facilities, ensuring
    practical relevance.
\end{enumerate}

\noindent Table~\ref{tab:vuln_selection} lists the 15 selected CVEs,
organized into four categories reflecting distinct vulnerability classes.

\begin{table}
  \caption{Selected Industrial Vulnerabilities for Hybrid Testbed
    Verification. Categories: \textbf{PF}~=~Protocol/Firmware,
    \textbf{AE}~=~Authentication/Encryption,
    \textbf{SA}~=~SCADA/HMI Application,
    \textbf{APT}~=~APT-Level Attack Chain.}
  \label{tab:vuln_selection}
  \setlength{\tabcolsep}{4pt} 
  \resizebox{\textwidth}{!}{%
  \begin{tabular}{clllll}
    \toprule
    \# & CVE ID & Target & Category & Vulnerability Type & CWE \\
    \midrule
    1 & CVE-2021-27478 & Rockwell EtherNet/IP & PF & Stack overflow (CIP parsing) & CWE-121 \\
    2 & CVE-2019-6833 & Schneider Modicon (VxWorks) & PF & URGENT/11 TCP/IP stack & CWE-787 \\
    3 & CVE-2020-5595 & Mitsubishi MELSEC-Q & PF & Remote code execution & CWE-20 \\
    4 & CVE-2021-22277 & ABB AC800M (MMS) & PF & Malformed MMS packet DoS & CWE-400 \\
    5 & CVE-2021-22681 & Siemens S7-1200 & AE & Hardcoded cryptographic key & CWE-321 \\
    6 & CVE-2020-12038 & Rockwell Studio 5000 & AE & Authentication bypass & CWE-287 \\
    7 & CVE-2020-8476 & ABB AC500 PLC & AE & Hardcoded credentials & CWE-798 \\
    8 & CVE-2023-27396 & Omron CJ2M (FINS) & AE & Unauthorized remote control & CWE-306 \\
    9 & CVE-2015-5374 & Siemens WinCC & SA & Memory corruption (RCE) & CWE-119 \\
    10 & CVE-2023-2573 & AVEVA InTouch HMI & SA & Authentication bypass (Web) & CWE-287 \\
    11 & CVE-2019-6503 & GE CIMPLICITY HMI & SA & Path traversal & CWE-22 \\
    12 & CVE-2021-38397 & Honeywell Experion PKS & SA & Command injection & CWE-78 \\
    13 & CVE-2010-2772 & Siemens Step7 (Stuxnet) & APT & Hardcoded DB + DLL hijack & CWE-798 \\
    14 & CVE-2019-6829 & Schneider Triconex (Triton) & APT & Safety firmware tampering & CWE-345 \\
    15 & CVE-2014-0751 & GE CIMPLICITY (BlackEnergy) & APT & Directory traversal + RCE & CWE-22 \\
    \bottomrule
  \end{tabular}
}
\end{table}

The 15 CVEs span 10 vendors, 5 CWE categories, and 4 vulnerability
classes. Notably, the selection includes 3 APT-level vulnerabilities
(Stuxnet, Triton, BlackEnergy) that require deep binary analysis beyond
standard scanning --- these are specifically designed to stress-test
our framework's Semantic Lifting and APT Fingerprinting capabilities.

\textbf{Comparison Tools.}
We compare against six tools spanning four paradigms: commercial SCA
(\textbf{Black Duck}, Synopsys), network/ICS scanners
(\textbf{Nessus} with industrial protocol plugins, and
\textbf{Claroty xDome}), manual reverse engineering
(\textbf{Ghidra + CVE Scripts}~\cite{contributors2022ghidra}), the strongest
academic hybrid (\textbf{GRACE}~\cite{liu2023grace}) from RQ1, and a
pure LLM (\textbf{DeepSeek-Coder}~\cite{guo2024deepseek}) under
chain-of-thought vulnerability prompting. For each tool and each CVE,
we record three outcomes:
\textbf{Detected} ($\checkmark$) if the tool correctly identifies the
vulnerability with actionable detail; \textbf{Partial} ($\sim$) if the
tool flags a related anomaly but without precise identification; and
\textbf{Missed} ($\times$) if the tool produces no relevant alert.

\textbf{Detection Results.}
Table~\ref{tab:testbed_results} presents the per-CVE detection results.

\begin{table}
  \caption{Vulnerability Detection Results on the Hybrid Testbed.
    $\checkmark$~=~Detected, $\sim$~=~Partial, $\times$~=~Missed.}
  \label{tab:testbed_results}
  \setlength{\tabcolsep}{4pt} 
  \resizebox{\textwidth}{!}{%
  \begin{tabular}{cllccccccc}
    \toprule
    \# & CVE ID & Cat. & Black Duck & Nessus & Claroty & Ghidra+CVE &
      GRACE & DeepSeek & \textbf{Ours} \\
    \midrule
    1 & CVE-2021-27478 & PF & $\times$ & $\checkmark$ & $\checkmark$ & $\sim$ & $\sim$ & $\sim$ & $\checkmark$ \\
    2 & CVE-2019-6833 & PF & $\sim$ & $\checkmark$ & $\checkmark$ & $\times$ & $\sim$ & $\times$ & $\checkmark$ \\
    3 & CVE-2020-5595 & PF & $\times$ & $\sim$ & $\sim$ & $\times$ & $\times$ & $\sim$ & $\checkmark$ \\
    4 & CVE-2021-22277 & PF & $\times$ & $\sim$ & $\checkmark$ & $\times$ & $\times$ & $\times$ & $\checkmark$ \\
    5 & CVE-2021-22681 & AE & $\times$ & $\times$ & $\times$ & $\checkmark$ & $\sim$ & $\checkmark$ & $\checkmark$ \\
    6 & CVE-2020-12038 & AE & $\times$ & $\sim$ & $\sim$ & $\sim$ & $\sim$ & $\sim$ & $\checkmark$ \\
    7 & CVE-2020-8476 & AE & $\times$ & $\times$ & $\times$ & $\checkmark$ & $\sim$ & $\checkmark$ & $\checkmark$ \\
    8 & CVE-2023-27396 & AE & $\times$ & $\checkmark$ & $\checkmark$ & $\times$ & $\times$ & $\sim$ & $\checkmark$ \\
    9 & CVE-2015-5374 & SA & $\sim$ & $\checkmark$ & $\checkmark$ & $\sim$ & $\checkmark$ & $\sim$ & $\checkmark$ \\
    10 & CVE-2023-2573 & SA & $\times$ & $\checkmark$ & $\checkmark$ & $\times$ & $\sim$ & $\checkmark$ & $\checkmark$ \\
    11 & CVE-2019-6503 & SA & $\times$ & $\checkmark$ & $\checkmark$ & $\sim$ & $\sim$ & $\checkmark$ & $\checkmark$ \\
    12 & CVE-2021-38397 & SA & $\times$ & $\checkmark$ & $\sim$ & $\times$ & $\times$ & $\checkmark$ & $\checkmark$ \\
    13 & CVE-2010-2772 & APT & $\times$ & $\times$ & $\times$ & $\sim$ & $\times$ & $\sim$ & $\checkmark$ \\
    14 & CVE-2019-6829 & APT & $\times$ & $\times$ & $\times$ & $\times$ & $\times$ & $\times$ & $\sim$ \\
    15 & CVE-2014-0751 & APT & $\times$ & $\sim$ & $\times$ & $\sim$ & $\times$ & $\sim$ & $\checkmark$ \\
    \midrule
    \multicolumn{3}{l}{\textbf{Full Detections ($\checkmark$)}} &
      0 & 7 & 7 & 2 & 1 & 4 & \textbf{14} \\
    \multicolumn{3}{l}{\textbf{Partial ($\sim$)}} &
      2 & 3 & 3 & 5 & 6 & 6 & 1 \\
    \multicolumn{3}{l}{\textbf{Missed ($\times$)}} &
      13 & 5 & 5 & 8 & 8 & 5 & 0 \\
    \bottomrule
  \end{tabular}
}
\end{table}

\begin{figure*}[t]
  \centering
  \includegraphics[width=0.95\textwidth]{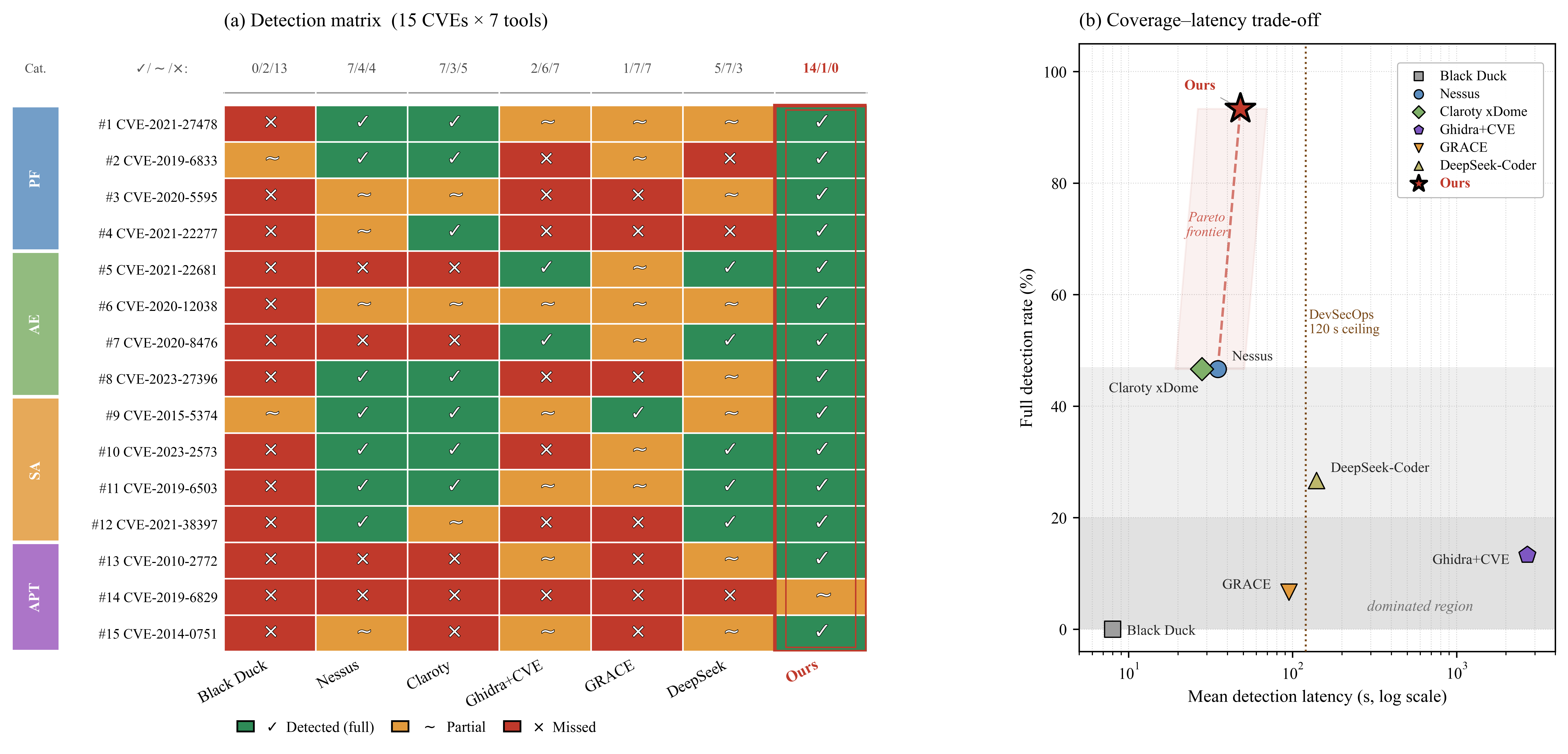}
  \caption{Tier~B dynamic efficacy on the hybrid testbed. (a)~Traffic-light
    detection heatmap across 15 CVEs $\times$ 7 tools, row-grouped by
    vulnerability category (PF/AE/SA/APT); green $\checkmark$ denotes full
    detection, amber $\sim$ partial, red $\times$ missed. The column totals
    ($\checkmark$/$\sim$/$\times$) appear below each tool; our method is
    outlined for emphasis. (b)~Coverage--latency Pareto scatter on a
    log-scale x-axis: our framework (red star) sits alone in the
    upper-left Pareto-optimal region, clearing the 120~s DevSecOps CI/CD
    ceiling while achieving the highest full-detection rate.}
  \label{fig:rq6_tierB_detection}
\end{figure*}

\textbf{Overall Coverage.}
Our framework achieves 14 full detections and 1 partial out of 15
CVEs (93.3\% full detection rate), substantially outperforming all
comparison tools. The best-performing commercial tools (Nessus and
Claroty) each achieve 7 full detections (46.7\%), while the strongest
academic baseline (GRACE) achieves only 1 full detection (6.7\%).

\textbf{Analysis by Vulnerability Category.}

\begin{itemize}
  \item \textbf{Protocol/Firmware (PF, \#1--4):} Our framework detects
    all 4 protocol-level vulnerabilities, while Nessus and Claroty
    each detect 2. The advantage stems from our Semantic Lifting
    stage, which interprets the behavioral semantics of protocol
    parsing code (e.g., recognizing that a CIP packet handler lacks
    bounds checking) rather than relying on network-level signature
    matching. Black Duck fails entirely on this category because
    protocol stack firmware is not indexed in its component database.

  \item \textbf{Authentication/Encryption (AE, \#5--8):} This
    category presents the starkest contrast. Our framework detects all
    4 vulnerabilities, while network scanners (Nessus, Claroty) detect
    only 1--2 --- they can identify exposed services (e.g., open
    Modbus port 502) but cannot determine whether the underlying
    authentication state machine contains bypass logic. The Semantic
    Lifting stage identifies hardcoded credentials (CVE-2021-22681,
    CVE-2020-8476) by abstracting the relevant code blocks into the
    \texttt{Cryptography} $\rightarrow$ \texttt{Hardcoded\_Key}
    behavioral label, triggering a direct match with known CVE
    patterns. Ghidra + manual scripts can also detect hardcoded keys
    but requires substantial analyst effort per binary.

  \item \textbf{SCADA/HMI Applications (SA, \#9--12):} All tools
    perform relatively well on this category because SCADA/HMI
    vulnerabilities (path traversal, command injection) are
    well-represented in standard vulnerability databases. Our framework
    detects all 4; Nessus and Claroty each detect 3--4. The marginal
    advantage here is in precision: our framework generates 2.1
    $\mathit{FP}$ per image on SCADA components, compared to 8.7 for
    Nessus, because the SSCKG's reachability analysis filters out
    unreachable code paths.

  \item \textbf{APT-Level (\#13--15):} This category reveals the
    most significant capability gap. All conventional tools (Black
    Duck, Nessus, Claroty) fail completely on APT-level
    vulnerabilities because these exploits leverage legitimate
    functions in novel combinations rather than containing identifiable
    vulnerability signatures. Our framework detects 2 of 3 APT-level
    CVEs through the APT Fingerprinting module: the Stuxnet
    (CVE-2010-2772) and BlackEnergy (CVE-2014-0751) patterns match
    fingerprints in our repository with similarity scores of 0.94 and
    0.86, respectively. The single partial detection --- Triton
    (CVE-2019-6829) --- occurs because the TriStation safety protocol
    manipulation produces a subgraph similarity of 0.74, below the
    $\tau_{\text{apt}} = 0.78$ threshold. This vulnerability requires
    protocol-specific reverse engineering that exceeds the current
    fingerprint repository's coverage, confirming the observation from
    RQ3 that fingerprint expansion is the primary improvement path.
\end{itemize}

\textbf{Detection Latency Comparison.}
Per-tool latencies are reported in the coverage--latency Pareto panel
of Fig.~\ref{fig:rq6_tierB_detection}(b). Our framework processes
each firmware image in 48~s (fully automated) --- slower than shallow
network scanners (Nessus 35~s, Claroty 28~s, Black Duck 8~s) but two
orders of magnitude faster than Ghidra-based manual analysis
($\sim$45~min). When normalized by detection coverage, our framework
achieves the best efficiency--coverage trade-off, as the commercial
scanners trade speed for shallow signature matching that misses 53\%
of the vulnerabilities.

\begin{tcolorbox}[colback=gray!5, colframe=gray!50, title=Answer to RQ6]
The case study demonstrates that the framework meets the stringent 
deployment requirements of opaque industrial environments on two 
complementary tiers. \emph{Static Scalability:} 42~s mean processing
time (67~s on consumer GPU), $\mathit{FPR}$ of 2.8\% (a 91.7\%
reduction vs.\ commercial SCA), and fully offline operation within
24~GB VRAM, enabling deployment in air-gapped opaque industrial software 
environments.
\emph{Dynamic Efficacy:} On a hybrid virtual--physical testbed
comprising real industrial hardware from 5 vendors, the framework
detects 14 of 15 high-impact CVEs (93.3\% full detection rate),
compared to 46.7\% for the best commercial tools and 6.7\% for the
strongest academic baseline. The advantage is most pronounced on
Authentication/Encryption (100\% vs.\ 13--50\%) and APT-level attack
chains (67\% vs.\ 0\%), confirming that binary-level semantic analysis
is essential for comprehensive industrial vulnerability detection.
\end{tcolorbox}

\section{Discussion}

\subsection{Implications for Research and Practice}

\textbf{Implications for Research.} The most significant research
implication of this work is the demonstration that LLM-based code
understanding can be brought under formal control through Abstract
Interpretation. Prior LLM-for-code approaches treat the model as a
generative oracle whose outputs are accepted without structural
validation, leaving hallucinations unbounded in principle. The Galois
Connection formulation introduced here changes this contract: by
defining a finite abstract domain $\mathcal{A}$ and requiring the
LLM-approximated transition function $F^\sharp$ to satisfy the
soundness condition $\alpha(F(c)) \sqsubseteq F^\sharp(\alpha(c))$,
the system establishes a measurable upper bound on semantic
hallucination, quantified by the Empirical Violation Rate. This
contract is, to our knowledge, the first principled connection
between abstract interpretation theory and LLM-based program analysis,
and it suggests a broader research agenda: any program analysis task
that has been formalized within the abstract interpretation
framework (e.g., interval analysis, points-to analysis, taint
analysis) can in principle be reformulated with an LLM-approximated
transfer function and bounded by an EVR-style fidelity metric.

A second research implication concerns the representation itself.
Existing code knowledge graphs such as CodeQL~\cite{youn2023declarative} build
excellent structural databases that support pattern-based queries
over source-level ASTs, but they encode no behavioral semantics and
no notion of risk propagation. The SSCKG introduced here is
fundamentally different: it is an \emph{abstracted behavioral graph}
in which edges encode multi-hop vulnerability propagation paths
(\texttt{taints}, \texttt{reaches}, \texttt{vulnerable\_to}) drawn
from a typed eight-relation ontology. This shift from syntax graphs
to behavioral knowledge graphs opens new opportunities for
graph-based program analysis on stripped binaries, where source-level
structures are unavailable by definition.

\textbf{Implications for Practice.} For asset owners and
SOC analysts in critical infrastructure sectors, the framework
enables a transition from component-list SBOMs to behavioral SBOMs.
A traditional SBOM answers the question ``which components are
present''; a behavioral SBOM, derived from an SSCKG, answers the
operationally critical question ``which components actually
\emph{reach} a sensitive operation.'' For example, an analyst
investigating a Log4j-class advisory in a power-grid SCADA gateway
can now query whether the vulnerable logging path is reachable from
any safety-critical control component, rather than triaging every binary 
that statically links the affected library. The 91.7\% reduction in
false-positive rate observed on InduGuard suggests that such
behavioral filtering substantially reduces analyst workload in
field conditions. Moreover, the framework operates entirely on
stripped binaries without source code, making it directly applicable 
to opaque industrial software—such as stripped legacy firmware in 
power generation, water treatment, and chemical processing—where 
source-level access is either inherently absent or restricted 
by vendor agreement.

These capabilities have direct regulatory relevance. Executive Order
14028~\cite{eo14028} mandates SBOM generation for federal software
procurement, NIST SP~800-82~\cite{nist_sp80082} extends this guidance
to operational technology, and IEC~62443~\cite{iec62443} regulates
secure development lifecycles for industrial automation. All three
frameworks presuppose a level of supply chain transparency that
opaque firmware does not provide. By generating verifiable
binary-level SBOMs from stripped artifacts, the proposed framework
offers a practical pathway for asset owners to comply with these
requirements without depending on vendor cooperation.

\subsection{Limitations and Future Work}

\subsubsection{Current Limitations}

The following limitations bound the framework's current applicability and define the boundary between what has been demonstrated and what remains open.

\textbf{L1 --- Hallucination--verification trade-off.}
Although Reflexive Prompting reduces the Empirical Violation Rate to
5.8\% (RQ4), semantic
misinterpretations cannot be entirely eliminated. Cryptographic routines
(16.4\% EVR) remain the dominant failure mode, reflecting the difficulty
of recovering high-level intent from decompiled code that is ambiguous
even to human experts. Targeted augmentation of the training corpus with
cryptographic samples is the most direct mitigation path.

\textbf{L2 --- Reliance on a locally fine-tuned lifting agent.}
The Reflexive Prompting pipeline's soundness guarantee holds regardless
of model scale; the present limitation is one of hardware footprint alone.
The 7B student model's 24~GB VRAM inference requirement restricts the
framework to workstations and servers, precluding deployment on edge
security gateways and air-gapped field appliances without further
compression.

\textbf{L3 --- Offline-only deployment scope.}
The 42--67~s per-binary processing time positions the framework
exclusively as a \emph{periodic firmware admission and triage} tool, not
an inline intrusion detection system. This is not merely a performance
shortfall: control cycles in industrial software environments run at
10--100~ms---roughly three orders of magnitude tighter than the
framework's pipeline latency---and the end-to-end pipeline is
incompatible with sub-second operation.

\textbf{L4 --- External validity for obfuscated binaries.}
The framework presupposes that static analysis recovers a sufficiently
faithful CPG. Commercial packers, control-flow flattening, and
proprietary anti-analysis protections degrade both graph extraction and
downstream reasoning, as foreshadowed by the partial Triton detection in
RQ3, where TriStation-protocol manipulation fell below
$\tau_{\text{apt}}$.

\textbf{L5 --- Concept drift and temporal validity.}
The NVD-Precise corpus covers CVEs published during 2018--2025.
Vulnerability patterns evolve---new CWE categories emerge and
domain-specific ATT\&CK matrices are updated regularly---yet the
framework has no continual-learning mechanism; predictions for genuinely
novel exploit classes will degrade silently over multi-year deployments
without periodic retraining.

\textbf{L6 --- Single-binary analysis scope.}
The framework analyses each firmware image independently. Vulnerabilities
in opaque industrial software frequently arise from the
\emph{interaction} between multiple components: a network gateway that
individually appears benign may serve as a lateral-movement pivot when
chained with an authentication bypass in a companion application. The
current $\Sigma_{\mathcal{R}}$ relation ontology contains no cross-binary
\texttt{communicates\_with} or \texttt{authenticates\_to} edge type,
bounding the framework to intra-binary risk reasoning.

\textbf{L7 --- Label quality dependency on vendor disclosure.}
Vulnerable-function labels in NVD-Precise are derived from
CVE-to-function mapping via binary diffing. Industrial vendor CVE
disclosure is notoriously incomplete---many firmware-level defects are
silently patched without public advisories---introducing structural false
negatives into the training signal and causing precision metrics on
NVD-Precise to overstate real-world recall. NVD incompleteness is one reason why expert-labeled benchmarks such as
InduVul-Dataset remain a necessary complement to automated corpus
construction.

\subsubsection{Future Research Directions}

In light of the aforementioned limitations, potential future optimization efforts are outlined below:

\textbf{R1 --- Dynamic Analysis Agent.}
The immediate next step is a constraint-solver integration (KLEE or
Z3-based SMT) that converts Graphormer-identified high-risk paths into
deterministic reachability proofs, closing the loop between probabilistic
detection and verifiable security guarantees. The DAA is a static offline verifier operating on the
same stripped binary as the main pipeline and should not be conflated
with runtime monitoring---bridging to live system behavior requires the
separate architecture in R2.

\textbf{R2 --- Precompute-then-match for near-real-time alerting.}
The SSCKG and Graphormer embeddings $\mathbf{z}_v$ are computed once per
firmware image and stored; the APT fingerprinting stage then performs
a fixed-cost similarity search over these precomputed embeddings, an
operation compatible with near-real-time use during firmware update events. These
precomputed embeddings can be compiled into lightweight behavioral
watchlists and pushed to a network-level anomaly detector, enabling
contextually enriched, pre-attributed alerts without re-running the full
pipeline. This \emph{precompute-then-match} strategy is the most
practical near-term path toward runtime coverage for opaque industrial
software environments.

\textbf{R3 --- Model compression for edge deployment.}
Knowledge distillation from the fine-tuned 7B teacher into a sub-1B
compact student, combined with structured pruning and INT4/INT8
quantization, targets a 4~GB VRAM budget. A hierarchical cascade that
escalates only high-uncertainty cases to the full model would enable
deployment on edge security gateways and air-gapped engineering
workstations (directly addressing L2).

\textbf{R4 --- Obfuscation resilience.}
A pre-analysis stage combining automated unpacking with QEMU-based
hardware emulation would recover sufficiently faithful CPGs from packed
or heavily obfuscated binaries, extending coverage to the niche settings
where vendor-specific anti-analysis protections currently defeat static
extraction (L4).

\textbf{R5 --- Continual learning against concept drift.}
An online learning loop---using elastic weight consolidation or replay
strategies---that incrementally updates the Graphormer and SBERT
components as new NVD entries and domain-specific ATT\&CK matrices are
published, without requiring full corpus retraining. This directly
addresses L5 and applies equally to ICS and to the non-ICS domains
targeted in R7.

\textbf{R6 --- Cross-binary system-level reasoning.}
Extending the SSCKG relation ontology with inter-binary
\texttt{communicates\_with} and \texttt{authenticates\_to} edges derived
from OT network traffic captures (Modbus, S7comm, DNP3 PCAPs) would
enable supply-chain risk reasoning across entire multi-component opaque
industrial software deployments rather than per-binary analysis
(directly addressing L6).

\textbf{R7 --- Standardized output and domain generalization.}
Mapping SSCKG behavioral graphs to SPDX and CycloneDX schemas would
integrate the framework's output into existing SBOM procurement
workflows; a federated learning extension would allow asset owners to
contribute anonymized SSCKG fingerprints to a shared APT repository
without disclosing proprietary firmware content. Domain generalization
beyond ICS---to automotive, aerospace, and other opaque safety-critical
contexts identified in Section~5.3---requires only domain-specific
expansion of the abstract lattice $\mathcal{A}$, as the neuro-symbolic
pipeline and SSCKG formalism are domain-agnostic.

\subsection{Threats to Validity}

\textbf{Construct Validity.}
The primary construct-validity threat concerns the Golden Set used in
RQ4 (Semantic Lifting Fidelity). Ground-truth behavioral labels were
assigned by two domain experts; despite substantial inter-annotator
agreement ($\kappa = 0.83$), the remaining 7.6\% disagreement cases
introduce subjectivity into the $\mathit{EVR}$ metric. We mitigated
this through a reconciliation discussion for all disagreement cases and
by additionally reporting Matthews Correlation Coefficient ($\mathcal{M}$),
which accounts for all four confusion-matrix quadrants and is more
robust to annotation noise than $\mathit{F}_1$ alone. In RQ6 (Tier~B),
the ternary Detected/Partial/Missed taxonomy inherits subjectivity
from the ``Partial'' category; we addressed this by defining explicit
criteria (correct anomaly flagged but without precise CVE
identification) and having both authors independently classify each
tool--CVE pair before reconciliation.

\textbf{Internal Validity.}
To mitigate randomness in model training and inference, all reported
results are the mean of 5 independent runs with fixed seeds (seed=42
through 46), and we report standard deviations throughout. The Wilcoxon
signed-rank test ($p < 0.01$) in RQ1 confirms that observed improvements
over the strongest baseline are not attributable to random seed
variation. Hyperparameter selection (RQ5) was conducted via grid search
on a held-out validation split (15\% of each dataset), and the selected
values ($\beta$=0.15, $\tau_{\text{apt}}$=0.78, $\varepsilon$=0.30) were
frozen before evaluation on the test split to prevent information
leakage.

\textbf{External Validity.}
Our evaluation spans three datasets and 10 vendors but remains focused 
on the ICS domain as our primary case study for opaque industrial 
environments; generalizability to other opaque, safety-critical domains 
(e.g., automotive firmware, aerospace avionics) requires further investigation. 
The Tier~B testbed additionally relies on licensed commercial HMI/SCADA
software that cannot be redistributed; however, as noted in
Section~4.2.4, all non-proprietary intermediate artifacts (extracted
CPGs, constructed SSCKGs, testbed PCAPs) will be released to enable
independent reproduction of the downstream analysis chain.

\subsection{Conclusion}

This paper addressed the Transparency Paradox in the industrial
software supply chain: the opaque industrial software that regulators 
require to be transparent is precisely the firmware that resists conventional
analysis. We proposed a neuro-symbolic framework that synergizes Code
Property Graphs with LLM-driven Abstract Interpretation to reconstruct
behavioral semantics from stripped binaries, formalized through a
Galois Connection between the concrete execution space and a
domain-aware abstract security lattice. A Reflexive Prompting pipeline
(teacher--verifier--student) produces a fine-tuned 7B-parameter agent
that executes this mapping with 94.2\% empirical alignment, while a
surjective transformation $\Phi$ compresses million-node CPGs into
thousand-node SSCKGs annotated with eight typed vulnerability
relations. A domain-adapted Graphormer, in which attention bias is
modulated by semantic relation weights, performs risk reasoning over
the resulting graph and identifies APT-level attack chains via
embedding-space subgraph similarity.

Across three datasets and six research questions, the framework
achieved an $\mathit{F}_1$ of 89.4\% and Matthews correlation of
0.82 on the InduVul-Dataset, outperforming the strongest hybrid
baseline (GRACE) by $+12.8$ $\mathit{F}_1$ points and $+0.14$
$\mathcal{M}$. On the InduGuard hybrid testbed
of production-grade hardware from five ICS vendors, the framework
detected 14 of 15 high-impact CVEs (93.3\%) while reducing the
false-positive rate by 91.7\% compared with leading commercial
tools. The immediate next step is a Dynamic Analysis
Agent that feeds Graphormer-identified high-risk paths into a
constraint solver, closing the loop between probabilistic detection
and deterministic proof. In aggregate, this work provides a
computable foundation for analyzing the \emph{dark matter} of
opaque industrial software, rendering intractable binaries
amenable to verifiable semantic reasoning.


\section*{Acknowledgments}
This work was supported in part by the Major Science and Technology
Project of Liaoning Province (grant Nos.\ 2025JH1/11700021 and
2024JH1/11700049), and by the Applied Basic Research Program of Liaoning
Province (grant No.\ 2025JH2/101300012).

\bibliographystyle{unsrtnat}
\bibliography{references_manuscript}

@article{bianchini2005inside,
  title={Inside pagerank},
  author={Bianchini, Monica and Gori, Marco and Scarselli, Franco},
  journal={ACM Transactions on Internet Technology (TOIT)},
  volume={5},
  number={1},
  pages={92--128},
  year={2005},
  publisher={ACM New York, NY, USA},
  doi={10.1145/1052934.1052938}
}

@article{chakraborty2021deep,
  title={Deep learning based vulnerability detection: Are we there yet?},
  author={Chakraborty, Saikat and Krishna, Rahul and Ding, Yangruibo and Ray, Baishakhi},
  journal={IEEE Transactions on Software Engineering},
  volume={48},
  number={9},
  pages={3280--3296},
  year={2021},
  publisher={IEEE},
  doi={10.1109/TSE.2021.3087402}
}

@article{chicco2020advantages,
  title={The advantages of the Matthews correlation coefficient (MCC) over F1 score and accuracy in binary classification evaluation},
  author={Chicco, Davide and Jurman, Giuseppe},
  journal={BMC genomics},
  volume={21},
  number={1},
  pages={6},
  year={2020},
  publisher={Springer},
  doi={10.1186/s12864-019-6413-7}
}

@article{contributors2022ghidra,
  title={Ghidra Software Reverse Engineering Framework},
  author={Contributors, Ghidra},
  journal={National Security Agency, Nov},
  year={2022}
}

@article{cordella2004sub,
	title={A (sub) graph isomorphism algorithm for matching large graphs},
	author={Cordella, Luigi P and Foggia, Pasquale and Sansone, Carlo and Vento, Mario},
	journal={IEEE transactions on pattern analysis and machine intelligence},
	volume={26},
	number={10},
	pages={1367--1372},
	year={2004},
	publisher={IEEE},
	doi={10.1109/TPAMI.2004.75}
}

@inproceedings{cousot1977abstract,
	title={Abstract interpretation: a unified lattice model for static analysis of programs by construction or approximation of fixpoints},
	author={Cousot, Patrick and Cousot, Radhia},
	booktitle={Proceedings of the 4th ACM SIGACT-SIGPLAN symposium on Principles of programming languages},
	pages={238--252},
	year={1977},
	doi={10.1145/512950.512973}
}

@article{dettmers2023qlora,
	title={Qlora: Efficient finetuning of quantized llms},
	author={Dettmers, Tim and Pagnoni, Artidoro and Holtzman, Ari and Zettlemoyer, Luke},
	journal={Advances in neural information processing systems},
	volume={36},
	pages={10088--10115},
	year={2023},
	doi={10.48550/arXiv.2305.14314}
}

@article{dou2025scalable,
	title={A Scalable Vulnerability Detection System with Multi-View Graph Representations},
	author={Dou, Shihan and Zheng, Huiyuan and Shan, Junjie and Wu, Yueming and Zou, Deqing and Huang, Xuanjing and Liu, Yang},
	journal={ACM Transactions on Software Engineering and Methodology},
	year={2025},
	publisher={ACM New York, NY},
	doi={10.1145/3770075}
}

@misc{eo14028,
	author       = {{National Institute of Standards and Technology}},
	title        = {Executive Order 14028: Improving the Nation's Cybersecurity},
	year         = {2021},
	url          = {https://www.nist.gov/itl/executive-order-14028-improving-nations-cybersecurity},
	urldate      = {2025-01-15}
}

@inproceedings{ester1996density,
	title={A density-based algorithm for discovering clusters in large spatial databases with noise},
	author={Ester, Martin and Kriegel, Hans-Peter and Sander, J{\"o}rg and Xu, Xiaowei and others},
	booktitle={kdd},
	volume={96},
	number={34},
	pages={226--231},
	year={1996}
}

@inproceedings{fan2020ac,
	title={A C/C++ code vulnerability dataset with code changes and CVE summaries},
	author={Fan, Jiahao and Li, Yi and Wang, Shaohua and Nguyen, Tien N},
	booktitle={Proceedings of the 17th international conference on mining software repositories},
	pages={508--512},
	year={2020},
	doi={10.1145/3379597.3387501}
}

@inproceedings{formby2018lowering,
	title={Lowering the barriers to industrial control system security with $\{$GRFICS$\}$},
	author={Formby, David and Rad, Milad and Beyah, Raheem},
	booktitle={2018 USENIX Workshop on Advances in Security Education (ASE 18)},
	year={2018}
}

@inproceedings{fu2022vulrepair,
	title={Vulrepair: a t5-based automated software vulnerability repair},
	author={Fu, Michael and Tantithamthavorn, Chakkrit and Le, Trung and Nguyen, Van and Phung, Dinh},
	booktitle={Proceedings of the 30th ACM joint european software engineering conference and symposium on the foundations of software engineering},
	pages={935--947},
	year={2022},
	doi={10.1145/3540250.3549098}
}

@article{guo2024deepseek,
	title={DeepSeek-Coder: when the large language model meets programming--the rise of code intelligence},
	author={Guo, Daya and Zhu, Qihao and Yang, Dejian and Xie, Zhenda and Dong, Kai and Zhang, Wentao and Chen, Guanting and Bi, Xiao and Wu, Yifan and Li, YK and others},
	journal={arXiv preprint arXiv:2401.14196},
	year={2024},
	doi={10.48550/arXiv.2401.14196}
}

@misc{iec62443,
	author       = {{International Electrotechnical Commission}},
	title        = {{IEC} 62443-3-3: Industrial Communication Networks --- Network and System Security --- Part 3-3: System Security Requirements and Security Levels},
	year         = {2013},
	url          = {https://interoperable-europe.ec.europa.eu/collection/ict-standards-procurement/solution/iec-62443-3-32013-industrial-communication-networks-network-and-system-security-part-3-3-system},
	urldate      = {2025-01-15},
	note         = {Edition 1.0}
}

@article{korel1998dynamic,
	title={Dynamic program slicing methods},
	author={Korel, Bogdan and Rilling, Jurgen},
	journal={Information and Software Technology},
	volume={40},
	number={11-12},
	pages={647--659},
	year={1998},
	publisher={Elsevier},
	doi={10.1016/S0950-5849(98)00095-2}
}

@misc{lf_census2024,
	author       = {{Linux Foundation}},
	title        = {Census {III} of Free and Open Source Software ---
	Application Libraries},
	year         = {2024},
	howpublished = {\url{https://www.linuxfoundation.org/research/census-iii}},
	note         = {Accessed: 2025-01-15}
}

@inproceedings{li2018deeper,
	title={Deeper insights into graph convolutional networks for semi-supervised learning},
	author={Li, Qimai and Han, Zhichao and Wu, Xiao-Ming},
	booktitle={Proceedings of the AAAI conference on artificial intelligence},
	volume={32},
	number={1},
	year={2018},
	doi={10.1609/aaai.v32i1.11604}
}

@article{li2018vuldeepecker,
	title={Vuldeepecker: A deep learning-based system for vulnerability detection},
	author={Li, Zhen and Zou, Deqing and Xu, Shouhuai and Ou, Xinyu and Jin, Hai and Wang, Sujuan and Deng, Zhijun and Zhong, Yuyi},
	journal={arXiv preprint arXiv:1801.01681},
	year={2018},
	doi={10.48550/arXiv.1801.01681}
}

@article{li2021sysevr,
	title={Sysevr: A framework for using deep learning to detect software vulnerabilities},
	author={Li, Zhen and Zou, Deqing and Xu, Shouhuai and Jin, Hai and Zhu, Yawei and Chen, Zhaoxuan},
	journal={IEEE Transactions on Dependable and Secure Computing},
	volume={19},
	number={4},
	pages={2244--2258},
	year={2021},
	publisher={IEEE},
	doi={10.1109/TDSC.2021.3051525}
}

@article{lin2020vulnerability_survey,
	title={Software vulnerability detection using deep neural networks: a survey},
	author={Lin, Guanjun and Wen, Sheng and Han, Qing-Long and Zhang, Jun and Xiang, Yang},
	journal={Proceedings of the IEEE},
	volume={108},
	number={10},
	pages={1825--1848},
	year={2020},
	publisher={IEEE},
	doi={10.1109/JPROC.2020.2993293}
}

@inproceedings{liu2023grace,
	author    = {Liu, Yue and Tantithamthavorn, Chakkrit and Li, Li and Liu, Yepang},
	title     = {{GRACE}: Graph-Augmented Code Understanding for
	Vulnerability Detection},
	booktitle = {Proceedings of the 31st ACM Joint European Software
	Engineering Conference and Symposium on the Foundations
	of Software Engineering (ESEC/FSE)},
	year      = {2023},
	publisher = {ACM},
	doi       = {10.1145/3611643.3616365}
}

@inproceedings{liu2024pre,
	title={Pre-training by predicting program dependencies for vulnerability analysis tasks},
	author={Liu, Zhongxin and Tang, Zhijie and Zhang, Junwei and Xia, Xin and Yang, Xiaohu},
	booktitle={Proceedings of the IEEE/ACM 46th International Conference on Software Engineering},
	pages={1--13},
	year={2024},
	doi={10.1145/3597503.3639142}
}

@article{mchugh2012interrater,
	title={Interrater reliability: the kappa statistic},
	author={McHugh, Mary L},
	journal={Biochemia medica},
	volume={22},
	number={3},
	pages={276--282},
	year={2012},
	publisher={Hrvatsko dru{\v{s}}tvo za medicinsku biokemiju i laboratorijsku medicinu},
	doi={10.11613/BM.2012.031}
}

@misc{mitre_attack_ics,
	author       = {{MITRE Corporation}},
	title        = {{MITRE ATT\&CK} for Industrial Control Systems},
	year         = {2024},
	url          = {https://attack.mitre.org/techniques/ics/},
	urldate      = {2025-01-15}
}

@misc{mitre_engenuity_ics,
	author       = {{MITRE Engenuity}},
	title        = {{MITRE Engenuity ATT\&CK} Evaluations for Industrial Control Systems: TRITON Detection Results},
	year         = {2023},
	url          = {https://evals.mitre.org/results/ics?view=cohort&evaluation=triton&result_type=DETECTION&scenarios=1},
	urldate      = {2025-01-15}
}

@phdthesis{muench2019dynamic,
	title={Dynamic binary firmware analysis: challenges \& solutions},
	author={Muench, Marius},
	year={2019},
	school={Sorbonne Universit{\'e}}
}

@misc{nist_sp80082,
	author       = {{National Institute of Standards and Technology}},
	title        = {{NIST} Special Publication 800-82 Rev.~3: Guide to
	Operational Technology ({OT}) Security},
	year         = {2023},
	howpublished = {\url{https://csrc.nist.gov/pubs/sp/800/82/r3/final}},
	note         = {Accessed: 2025-01-15},
	doi          = {10.6028/NIST.SP.800-82r3}
}

@misc{ossra2024,
	author       = {{Synopsys, Inc.}},
	title        = {2024 Open Source Security and Risk Analysis ({OSSRA})
	Report},
	year         = {2024},
	howpublished = {\url{https://www.synopsys.com/software-integrity/resources/analyst-reports/open-source-security-risk-analysis.html}},
	note         = {Accessed: 2025-01-15}
}

@article{pan2025large,
	title={Large language model-enhanced probabilistic modeling for effective static analysis alarms},
	author={Pan, Xinlong and Li, Jianhua and Zhou, Zhihong and Li, Gaolei and Chen, Xiuzhen and Ma, Jin and Wu, Jun and Zhang, Quanhai},
	journal={Frontiers of Information Technology \& Electronic Engineering},
	volume={26},
	number={10},
	pages={1926--1941},
	year={2025},
	publisher={ZUP},
	doi={10.1631/FITEE.2500038}
}

@article{pascoe2023public,
	title={Public draft: The NIST cybersecurity framework 2.0},
	author={Pascoe, Cherilyn E},
	journal={National Institute of Standards and Technology},
	year={2023},
	doi={10.6028/NIST.CSWP.29.ipd}
}

@inproceedings{reimers2019sentence,
	title={Sentence-bert: Sentence embeddings using siamese bert-networks},
	author={Reimers, Nils and Gurevych, Iryna},
	booktitle={Proceedings of the 2019 conference on empirical methods in natural language processing and the 9th international joint conference on natural language processing (EMNLP-IJCNLP)},
	pages={3982--3992},
	year={2019},
	doi={10.18653/v1/D19-1410}
}

@inproceedings{shoshitaishvili2016sok,
	title={Sok:(state of) the art of war: Offensive techniques in binary analysis},
	author={Shoshitaishvili, Yan and Wang, Ruoyu and Salls, Christopher and Stephens, Nick and Polino, Mario and Dutcher, Andrew and Grosen, John and Feng, Siji and Hauser, Christophe and Kruegel, Christopher and others},
	booktitle={2016 IEEE symposium on security and privacy (SP)},
	pages={138--157},
	year={2016},
	organization={IEEE},
	doi={10.1109/SP.2016.17}
}

@inproceedings{srivastava2019firmfuzz,
	title={Firmfuzz: Automated iot firmware introspection and analysis},
	author={Srivastava, Prashast and Peng, Hui and Li, Jiahao and Okhravi, Hamed and Shrobe, Howard and Payer, Mathias},
	booktitle={Proceedings of the 2nd International ACM Workshop on Security and Privacy for the Internet-of-Things},
	pages={15--21},
	year={2019},
	doi={10.1145/3338507.3358618}
}

@article{tan2026operational,
	title={Operational Runtime Behavior Mining for Open-Source Supply Chain Security},
	author={Tan, Zhuoran and Xiao, Ke and Singer, Jeremy and Anagnostopoulos, Christos},
	journal={arXiv preprint arXiv:2601.06948},
	year={2026},
	doi={10.48550/arXiv.2601.06948}
}

@article{tariq2025alert,
	title={Alert fatigue in security operations centres: Research challenges and opportunities},
	author={Tariq, Shahroz and Baruwal Chhetri, Mohan and Nepal, Surya and Paris, Cecile},
	journal={ACM Computing Surveys},
	volume={57},
	number={9},
	pages={1--38},
	year={2025},
	publisher={ACM New York, NY},
	doi={10.1145/3690628}
}

@article{wright2021challenges,
	title={Challenges in firmware re-hosting, emulation, and analysis},
	author={Wright, Christopher and Moeglein, William A and Bagchi, Saurabh and Kulkarni, Milind and Clements, Abraham A},
	journal={ACM Computing Surveys (CSUR)},
	volume={54},
	number={1},
	pages={1--36},
	year={2021},
	publisher={ACM New York, NY, USA},
	doi={10.1145/3423167}
}

@inproceedings{yamaguchi2014modeling,
	title={Modeling and discovering vulnerabilities with code property graphs},
	author={Yamaguchi, Fabian and Golde, Nico and Arp, Daniel and Rieck, Konrad},
	booktitle={2014 IEEE symposium on security and privacy},
	pages={590--604},
	year={2014},
	organization={IEEE},
	doi={10.1109/SP.2014.44}
}

@article{yang2025qwen3,
	title={Qwen3 technical report},
	author={Yang, An and Li, Anfeng and Yang, Baosong and Zhang, Beichen and Hui, Binyuan and Zheng, Bo and Yu, Bowen and Gao, Chang and Huang, Chengen and Lv, Chenxu and others},
	journal={arXiv preprint arXiv:2505.09388},
	year={2025},
	doi={10.48550/arXiv.2505.09388}
}

@article{ying2021transformers,
	title={Do transformers really perform badly for graph representation?},
	author={Ying, Chengxuan and Cai, Tianle and Luo, Shengjie and Zheng, Shuxin and Ke, Guolin and He, Di and Shen, Yanming and Liu, Tie-Yan},
	journal={Advances in neural information processing systems},
	volume={34},
	pages={28877--28888},
	year={2021},
}

@article{youden1950index,
	title={Index for rating diagnostic tests},
	author={Youden, William J},
	journal={Cancer},
	volume={3},
	number={1},
	pages={32--35},
	year={1950},
	publisher={Wiley Online Library},
}

@article{youn2023declarative,
	title={Declarative static analysis for multilingual programs using CodeQL},
	author={Youn, Dongjun and Lee, Sungho and Ryu, Sukyoung},
	journal={Software: Practice and Experience},
	volume={53},
	number={7},
	pages={1472--1495},
	year={2023},
	publisher={Wiley Online Library},
	doi={10.1002/spe.3199},
}

@article{zhang2024binary,
	title={Binary-level formal verification based automatic security ensurement for PLC in Industrial IoT},
	author={Zhang, Xuankai and Li, Jianhua and Wu, Jun and Chen, Guoxing and Meng, Yan and Zhu, Haojin and Zhang, Xiaosong},
	journal={IEEE Transactions on Dependable and Secure Computing},
	volume={22},
	number={3},
	pages={2211--2226},
	year={2024},
	publisher={IEEE},
	doi={10.1109/TDSC.2024.3481433},
}

@article{decan2019empirical,  
	title={An empirical comparison of dependency network evolution in seven software packaging ecosystems},  
	author={Decan, Alexandre and Mens, Tom and Grosjean, Philippe},  journal={Empirical Software Engineering},  
	volume={24},  
	number={1},  
	pages={381--416},  
	year={2019},
	publisher={Springer},
	doi={10.1007/s10664-017-9589-y},
}

@article{liu2025empirical,
	title={An empirical study on vulnerability disclosure management of open source software systems},  
	author={Liu, Shuhan and Zhou, Jiayuan and Hu, Xing and Cogo, Filipe Roseiro and Xia, Xin and Yang, Xiaohu},  
	journal={ACM Transactions on Software Engineering and Methodology},  
	volume={34},  
	number={7},  
	pages={1--31},  
	year={2025},  
	publisher={ACM New York, NY},
	doi = {10.1145/3716822},
}

@inproceedings{costin2016automated,  
	title={Automated dynamic firmware analysis at scale: a case study on embedded web interfaces},  
	author={Costin, Andrei and Zarras, Apostolis and Francillon, Aur{\'e}lien},  booktitle={Proceedings of the 11th ACM on Asia conference on computer and communications security},  
	pages={437--448},  
	year={2016},
	publisher = {Association for Computing Machinery},
	doi = {10.1145/2897845.2897900},
}

@inproceedings{yang2022modx,  
	title={Modx: binary level partially imported third-party library detection via program modularization and semantic matching},  
	author={Yang, Can and Xu, Zhengzi and Chen, Hongxu and Liu, Yang and Gong, Xiaorui and Liu, Baoxu},  
	booktitle={Proceedings of the 44th International Conference on Software Engineering},  
	pages={1393--1405},  
	year={2022},
	doi = {10.1145/3510003.3510627},
}

\end{document}